\documentstyle[epsf,aps,prb,preprint]{revtex}
\tightenlines
\begin{document}
\draft
\widetext
\title{Paramagnetic State of Degenerate Double Exchange Model: \\
a Non-Local CPA Approach}
\author{I. V. Solovyev}
\address{JRCAT-Angstrom Technology Partnership, c/o AIST, \\
Central 4,
1-1-1 Higashi, Tsukuba, Ibaraki 305-0046, Japan\\
and \\
Institute of Metal Physics, Russian Academy of Sciences, \\
Ekaterinburg GSP-170, Russia}
\date{\today}
\maketitle

\widetext
\begin{abstract}
The off-diagonal disorder caused by random spin orientations
in the paramagnetic (PM) state of the double exchange (DE) model
is described by using
the coherent-potential-approximation (CPA), which is
combined with the variational mean-field approach for the Curie
temperature ($T_C$). Our CPA approach is essentially non-local and
based on the perturbation theory expansion for the $T$-matrix with
respect to the fluctuations of hoppings from
the
''mean values'' specified by matrix elements of the self energy,
so that in the first order it becomes identical
to the DE theory by de Gennes.
The second-order effects, considered in the present work, can be viewed
as an extension of this theory.
They are not negligible, and lead to a substantial reduction
of $T_C$ in the one-orbital case.
Even more dramatic changes are expected in the case of orbital
degeneracy, when each site of the cubic lattice
is represented by
two $e_g$ orbitals, which also specify the form of interatomic transfer
integrals.
Particularly,
the existence of two Van Hove singularities in the spectrum of degenerate DE model
(one of which is expected near the Fermi level
in the 30\%-doped LaMnO$_3$)
may lead to the branching of CPA solutions,
when the Green function and the self energy become
double-valued functions in certain region of the complex plane.
Such a behavior can be interpreted
as an intrinsic inhomogeneity of the PM state, which consists of
two
phases characterized by different electronic densities.
The phase separation occurs below certain transition temperature,
$T_P$, and naturally explains
the
appearance
of several magnetic transition points, which are frequently seen in
manganites.
We discuss possible implications of our theory to the experimental situation
in manganites, as well as possible extensions which needs to be done
in order to clarify its credibility.

\end{abstract}



\section{Introduction}
\label{sec:intr}

  The nature of paramagnetic (PM) state in perovskite
manganese oxides (the manganites)
is one of the fundamental questions, the answer to which
is directly related with
understanding of the phenomenon of colossal magnetoresistance.

  There is no doubts that any theoretical model for manganites
should include
(at least, as one of the main ingredients)
the double exchange (DE) physics, which enforces the
atomic Hund's rule and penalizes the
hoppings of polarized
$e_g$ electrons to the sites with the opposite direction of
the
localized $t_{2g}$ spins.\cite{Zener,AndersonHasegawa,deGennes}
If the spins are treated classically, the corresponding
Hamiltonian
is given,
in the local coordinate frame specified by the
directions
${\bf e_i}$$=$$(\cos\phi_{\bf i} \sin\theta_{\bf i},
           \sin\phi_{\bf i} \sin\theta_{\bf i},
           \cos\theta_{\bf i})$
of the spin magnetic moments,
by\cite{MullerHartmann}
\begin{equation}
{\cal H}_{\bf ij} = - \xi_{\bf ij} t_{\bf ij},
\label{eqn:DE_Hamiltonian}
\end{equation}
where
$t_{\bf ij}$ are the bare transfer integrals between sites ${\bf i}$ and ${\bf j}$
(which can be either scalars or $2$$\times$$2$ matrices depending on how the
orbital degeneracy of the $e_g$ states is treated in the model),
and
$\xi_{\bf ij}$$=$$\cos \frac{\theta_{\bf i}}{2} \cos \frac{\theta_{\bf j}}{2}
$$+$$\sin \frac{\theta_{\bf i}}{2} \sin \frac{\theta_{\bf j}}{2}
e^{-i(\phi_{\bf i} - \phi_{\bf j})}$
describes their modulations caused by
the
deviation from the ferromagnetic (FM) alignment in the pair ${\bf i}$-${\bf j}$.

  The form of Hamiltonian (\ref{eqn:DE_Hamiltonian}) implies that the spin
magnetic moments are saturated (due to the strong Hund's rule
coupling) and the spin disorder corresponding to the PM state
is in fact an {\it orientational} spin disorder.
Despite an apparent simplicity of the DE Hamiltonian (\ref{eqn:DE_Hamiltonian}),
the description of this orientational spin disorder is not an easy task and so far
there were only
few theoretical model,
which were based on rather severe (and presumably unsatisfactory)
approximations.

  The first one was proposed by de Gennes more than forty years ago.\cite{deGennes}
In his theory, all $\xi_{\bf ij}$ are replaced by an averaged value $\overline{\xi}$,
so that the spin disorder enters the model
only
as a renormalization (narrowing) of the
$e_g$-bandwidth. The effect is not particularly strong and the fully
disordered PM state corresponds to $\overline{\xi}$$=$$\frac{2}{3}$.
A generalization of this theory to the case of quantum spins was given by
Kubo and Ohata.\cite{KuboOhata}
The same idea was exploited recently in
a number of theories aiming to study the behavior of orbital degrees of freedom
at elevated temperature,\cite{orbitaltheories}
but based on the same kind of simplifications for the spin disorder and its
effects on the kinetic energy.

  Another direction, which features more recent activity, is the single-site
dynamical mean-field theory (DMFT, see Ref.~\onlinecite{DMFT} for a review)
for the FM Kondo lattice model.\cite{Furukawa,Chattopadhyay}
The model itself can be viewed as a prototype of the DE
Hamiltonian (\ref{eqn:DE_Hamiltonian}) before projecting out the minority-spin
states in the local coordinate frame.\cite{MullerHartmann}
If the localized $t_{2g}$ spins are treated classically (that is typically the case),
this method is similar to the disordered local moment approach proposed by
Gyorffy {\it et~al.},\cite{DLM}
and based on the coherent-potential-approximation (CPA) for the electronic structure
of the disordered state.\cite{Oguchi}

  Now it is almost generally accepted that both approaches are inadequate as they
fail to explain not even all, but a certain number of observations in
manganites of a principal character such as the absolute value and the doping
dependence of the Curie temperature ($T_C$),\cite{comment.6}
the insulating behavior above $T_C$,\cite{Urushibara} and a rich magnetic
phase diagram along the temperature axis,
which typically show a number of magnetic phase transitions\cite{Tomioka}
and the phase coexistence\cite{Uehara,Fath,Lynn} in certain temperature interval.
Therefore, it is clear that the theory must be revised.

  Then, there are two possible ways to precede.
One is to modernize the model itself by including additional ingredients
such as the Jahn-Teller distortion,\cite{Millis1}
the Coulomb correlations,\cite{orbitaltheories,Ishihara}
and the disorder effects caused by the chemical substitution.\cite{Moreo}
Another possibility is to stick to the basic concept of the DE physics
and try to formulate a more advanced theory of the spin disorder described by
the Hamiltonian (\ref{eqn:DE_Hamiltonian}), which would go beyond the simple
scaling theory by de Gennes\cite{deGennes} as well as the single-site approximation
inherent to DMFT.\cite{Furukawa,Chattopadhyay}

  An attention to the second direction was drawn recently by Varma.\cite{Varma}
The main challenge to the theoretical description of the orientational
spin disorder in the
DE systems comes from the fact that it enters the
Hamiltonian (\ref{eqn:DE_Hamiltonian})
as an {\it off-diagonal} disorder of interatomic transfer integrals,
which presents a serious and not well investigated problem.
In the present work we try to investigate some possibilities along this line
by employing a non-local CPA approach.

  What do we expect?

\begin{figure}
\centering
\noindent
\epsfxsize=10cm  \epsfbox{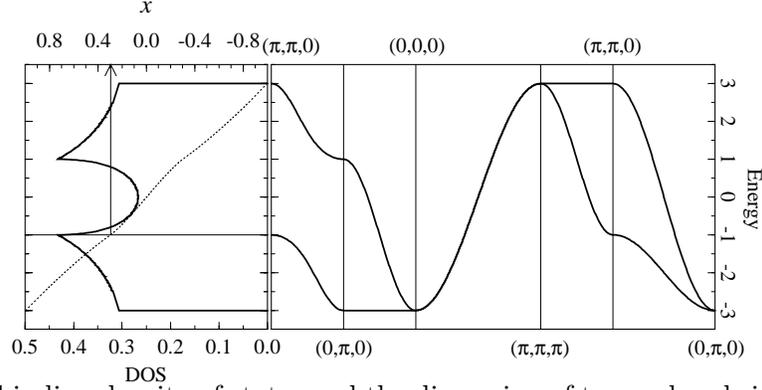}
\caption{Tight-binding density of states and the dispersion of two $e_g$ bands
         in the ferromagnetic state (in units of $dd\sigma$ transfer integral).
         The dotted line shows the positions of the Fermi level as a function
         of hole-concentration $x$,
         which is related with the integrated density of states as $x$$=$$1$$-$$n$.
         Note the existence of two Van Hove singularities at $(\pi,\pi,0)$ and
         $(0,\pi,0)$, responsible
         for the kinks
         of density of states at $\pm$1. The first singularity
         is located near the Fermi level
         when $x$$\simeq$$0.3$ (shown by arrow).}
\label{fig.FDOS}
\end{figure}

\begin{enumerate}
\item
It was realized very recently that many aspects of seemingly complicated
low-temperature behavior
of the doped manganites (the rich magnetic phase diagram,
optical properties, etc.) can be understood from the viewpoint of DE physics,
if the latter is considered in the combination with the realistic electronic structure
for the itinerant $e_g$ electrons and takes into account the strong dependence of this
electronic structure on the magnetic ordering
(see, e.g., Ref.~\onlinecite{Springer02} and references therein).
If this scenario is correct and can be extended to the high-temperature regime
(that is still a big question), there should be something peculiar in the
electronic structure of the disordered PM state, which can be linked
to the unique properties of manganites.
In order to gain insight into this problem, let us start with the DE picture
by de Gennes,
and consider it in the combination with
the correct form of the transfer integrals
between two $e_g$ orbitals on the cubic lattice.\cite{SlaterKoster}
Then, the properties of the PM state should be directly related with details
of the electronic structure of the FM state, which is shown in Fig.~\ref{fig.FDOS}
and connected with the PM electronic structure by the scaling transformation.
This electronic structure is indeed
very peculiar because of two Van Hove singularities
at the $(\pi,\pi,0)$ and $(0,\pi,0)$
points of the Brillouin zone, which are responsible for two kinks of density
of states at $\pm$$1$.
It is also interesting to note that the first singularity appears near the
Fermi surface when the hole concentration is close to $0.3$, i.e. in the most
interesting regime from the viewpoint of colossal magnetoresistance.\cite{Tokura}
Such a behavior was discussed by Dzero, Gor'kov and Kresin.\cite{Gorkov}
They also argued that this singularity can contribute to
the $T^{3/2}$ dependence of the specific heat in the FM state.
If so, what is the possible role of these singularities in the case of the
spin disorder?
Note that apart from the single-site approximation, all recent DMFT
calculations\cite{Furukawa,Chattopadhyay} employed a model semi-circular
density of states, and therefore could not address this problem.
\item
There are many anticipations in the literature that there is some hidden degree
of freedom which controls the properties of perovskite manganites.
A typical example is the picture of orbital disorder proposed in
Ref.~\onlinecite{Ishihara} in connection with the anomalous behavior of
the optical conductivity in the FM state of perovskite manganites.
According to this picture,
the large on-site Coulomb interaction gives rise to the orbital polarization
at each site of the system.
The local orbital polarizations
remain even in the cubic FM phase,
but without the long-range ordering.
In the present work we will show that, in principle, by considering non-local
effects in the framework of
pure DE model, one may have an alternative scenario,
when there is a certain degree of freedom which does control the properties
of manganites. However, contrary to the orbital polarization,
this parameter is
essentially non-local and attached to the bond of the
DE system rather than to the site.
\item
There were many debated about the phase separation in perovskite
manganites,\cite{Nagaev,Dagotto}
and according to some scenarios
this effect plays an important role also at elevated temperatures,
being actually the main trigger behind phenomenon of the colossal
magnetoresistance.\cite{Moreo}
The problem was intensively studied numerically, using the
Monte Carlo techniques.\cite{Moreo,Dagotto}
If this is indeed the case, what does it mean on the language of
{\it analytical} solutions of the DE model (or its refinements)?
Presumably, the only possibility to have two (and more) phases at the same
time is to admit that the self energy (and the Green function) of the model
can be a multi-valued function
in certain region of the complex plane.
Such a behavior of non-linear CPA equations was considered as one of the
main troublemakers in the past,\cite{NickelButler,Elliott}
but may have some physical explanation in the light of newly proposed
ideas of phase separation.
\end{enumerate}

  The rest of the paper is organized as follows. In Sec.~\ref{sec:thermal}
we briefly review the main ideas of the variational mean-field approach.
In Sec.~\ref{sec:cluster_CPA} we describe general ideas of the
non-local CPA to the problem of orientational
spin disorder in the DE systems.
In Sec.~\ref{sec:1band} we consider the CPA solution for the PM phase of the
one-orbital model and evaluate the Curie temperature.
We will argue that two seemingly different approaches to the problem of spin
disorder in the DE model, one of which was proposed by de Gennes\cite{deGennes}
and the other one is based on the DMFT\cite{Furukawa,Chattopadhyay}, have
a common basis and can be regarded as CPA-type approaches, but supplemented with
different types of approximations.
In Sec.~\ref{sec:2band}
we consider more realistic two-orbital case for the $e_g$ electrons and argue that
it is qualitatively different from the behavior of one-orbital model.
Particularly, the CPA self-energy becomes the double-valued function in certain
region of the complex plane, that can be related with an intrinsic inhomogeneity
of the PM state of the DE model.
In Sec.~\ref{sec:discussions} we summarize the main results of our work,
discuss possible connections with the experimental behavior of
perovskite manganites as well as possible extensions.

\section{Calculation of Thermal Averages}
\label{sec:thermal}

  In order to proceed with the finite temperature description of the
DE systems we adopt the variational mean-field approach.\cite{deGennes,KuboOhata,Alonso}
Namely, we assume that the thermal (or orientational)
average of any physical quantity is given in
terms of the single spin orientation distribution function, which depends only on the
angle between the local spin and an effective molecular field $\mbox{\boldmath$\lambda$}$,
\begin{equation}
p_{\bf i}({\bf e}_{\bf i}) \propto \exp\left(\mbox{\boldmath$\lambda$} \cdot {\bf e_i}\right),
\label{eqn:dfunction0}
\end{equation}
while all correlations between different spins are neglected.\cite{comment.13}
In the case of
PM-FM transition, the
effective field can be chosen as
$\mbox{\boldmath$\lambda$}$$=$$(0,0,\lambda)$.

  The first complication comes from the fact that the DE Hamiltonian
(\ref{eqn:DE_Hamiltonian}) is formulated in
the {\it local} coordinate frame,
in which the spin quantization axes at different sites are specified by the different
direction $\{ {\bf e_i} \}$.
Therefore, we should clarify the meaning of orientational averaging in the
local coordinate frame.\cite{comment.4}
From our point of view, it is logical that in order to calculate the thermal
averages associated with an arbitrary chosen site ${\bf 0}$, the global
coordinate frame should be specified by the direction
${\bf e_0}$$=$$(\cos\phi_{\bf 0} \sin\theta_{\bf 0},
           \sin\phi_{\bf 0} \sin\theta_{\bf 0},
           \cos\theta_{\bf 0})$,
so that at each instant the spin moment at the site ${\bf 0}$
is aligned along the $z$-direction.
The averaging over all possible directions ${\bf e_0}$ of the global coordinate frame
in the molecular field $\mbox{\boldmath$\lambda$}$ can be performed as the second step.

  Then, corresponding distribution function at the site ${\bf 0}$ is given by
Eq.~(\ref{eqn:dfunction0}). The distribution functions
at other sites can be constructed as follows.
The transformation to the coordinate frame associated with the site ${\bf 0}$ is
given by the matrix:
$$
\widehat{R}=
\left(
\begin{array}{ccc}
\cos^2 \phi_{\bf 0} (\cos\theta_{\bf 0}-1)+1 & \sin\phi_{\bf 0} \cos\phi_{\bf 0} (\cos\theta_{\bf 0}-1) &
-\cos\phi_{\bf 0} \sin\theta_{\bf 0} \\
\sin\phi_{\bf 0} \cos\phi_{\bf 0} (\cos\theta_{\bf 0}-1) & \sin^2 \phi_{\bf 0} (\cos\theta_{\bf 0}-1)+1 &
-\sin\phi_{\bf 0} \sin\theta_{\bf 0} \\
\cos\phi_{\bf 0} \sin\theta_{\bf 0} & \sin\phi_{\bf 0} \sin\theta_{\bf 0} & \cos\theta_{\bf 0} \\
\end{array}
\right).
$$
In the new coordinates, the ${\bf i}$-th moment has the direction
${\bf e}'_{\bf i}$$=$$\widehat{R} {\bf e}_{\bf i}$$\equiv$$
(\cos\phi'_{\bf i} \sin\theta'_{\bf i},
           \sin\phi'_{\bf i} \sin\theta'_{\bf i},
           \cos\theta'_{\bf i})$,
and the effective field becomes
$\mbox{\boldmath$\lambda$}'$$=$$\widehat{R}\mbox{\boldmath$\lambda$}$$=$$
(-\cos\phi_{\bf 0} \sin\theta_{\bf 0}, -\sin\phi_{\bf 0} \sin\theta_{\bf 0}, \cos\theta_{\bf 0}) \lambda$.

  Obviously, while $\mbox{\boldmath$\lambda$}'$ and ${\bf e}'_{\bf i}$ depend on
$\theta_{\bf 0}$ and $\phi_{\bf 0}$,
the new distribution function
$p_{\bf i}({\bf e}'_{\bf i})$$\propto$$\exp(\mbox{\boldmath$\lambda$}' \cdot {\bf e}'_{\bf i})$  --
does not, and
up to this stage the transformation
to the new coordinate frame was only the change of the notations.
In order to obtain the total distribution function
$P_{\bf i}({\bf e}'_{\bf i})$ for ${\bf i}$$\neq$$0$, formulated in the
{\it local coordinates of the site} ${\bf 0}$ {\it and taking into account the motion of}
${\bf e_0}$ in the molecular field $\mbox{\boldmath$\lambda$}$,
$p_{\bf i}({\bf e}'_{\bf i})$ should be averaged over ${\bf e_0}$ with the weight
$p_{\bf 0}({\bf e}_{\bf 0})$:
\begin{equation}
P_{\bf i}({\bf e}'_{\bf i},\lambda) = \frac{1}{\nu}
\int d \mbox{\boldmath$\Omega$}_{\bf 0}
\exp\left(\mbox{\boldmath$\lambda$}' \cdot {\bf e}'_{\bf i} +
\mbox{\boldmath$\lambda$} \cdot {\bf e}_{\bf 0}\right).
\label{eqn:dfunction1}
\end{equation}
The normalization constant $\nu$ is obtained from the condition:
$$
\int d \mbox{\boldmath$\Omega$}'_{\bf i} P_{\bf i}({\bf e}'_{\bf i},\lambda) = 1.
$$
The form of Eq.~(\ref{eqn:dfunction1}) implies that the directions of magnetic
moments are not correlated (in the spirit of the mean-field approach)
and the averaging over ${\bf e}_{\bf 0}$ can be performed independently for different
sites of the system.
For the analysis
the PM state and the magnetic
transition temperature,
it is sufficient to consider
the small-$\lambda$ limit.
Then, Eq.~(\ref{eqn:dfunction1}) becomes:
\begin{equation}
P_{\bf i}({\bf e}'_{\bf i},\lambda) \simeq \frac{1}{4\pi}
\left( 1 + \frac{1}{3} \cos\theta'_{\bf i} \lambda^2 \right).
\label{eqn:dfunction}
\end{equation}
The thermal average of the function $F({\bf e}'_{\bf i})$,
formulated in the coordinate frame of the site ${\bf 0}$ and taking into
account the motion of both ${\bf e_0}$ and ${\bf e_i}$,
is given by
$$
\overline{F}(\lambda)=
\int d \mbox{\boldmath$\Omega$}'_{\bf i} P_{\bf i}({\bf e}'_{\bf i},\lambda) F({\bf e}'_{\bf i}).
$$

  The spin entropy can be computed in terms of the molecular field $\lambda$ as:\cite{deGennes}
$$
-TS(\lambda) = k_B T \int d \mbox{\boldmath$\Omega$}'_{\bf i} P_{\bf i}({\bf e}'_{\bf i},\lambda)
\ln P_{\bf i}({\bf e}'_{\bf i},\lambda).
$$
In the second order of $\lambda$ this yields (both for ${\bf i}$$=$$0$ and
${\bf i}$$\neq$$0$):
\begin{equation}
-TS(\lambda) \simeq \frac{k_B T}{6} \lambda^2.
\label{eqn:entropy}
\end{equation}

  Then, the free energy of the DE model is given by\cite{deGennes,Alonso}
\begin{equation}
{\cal F}(T,\lambda) = E_D(T,\lambda) - TS(\lambda),
\label{eqn:fenergy}
\end{equation}
where $E_D(T,\lambda)$ is the electron free energy (or the double exchange energy):
\begin{equation}
E_D(T,\lambda) = - \int_{-\infty}^{+\infty}
dz f_T(z-\mu) \overline{n}(z,\lambda),
\label{eqn:DEenergy}
\end{equation}
calculated in terms of the (orientationally averaged) integrated
density of states $\overline{n}(z,\lambda)$.
$f_T(z$$-$$\mu)$$=$$[\exp(\frac{z-\mu}{k_BT})$$+$$1]^{-1}$
is the Fermi-Dirac function ($\mu$ being the
chemical potential).

  The best approximation for the molecular field
$\lambda$ is that which minimizes the free energy (\ref{eqn:fenergy}).
Assuming that the transition to the FM state is continuous (of the second order),\cite{comment.5}
$T_C$ can be found from equation:
\begin{equation}
\left. \frac{\partial^2 {\cal F}(T_C,\lambda)}{\partial\lambda^2}\right|_{\lambda = 0}=0.
\label{eqn:TC}
\end{equation}
In practice, the derivative
$\partial^2 E_D(T,\lambda) / \partial\lambda^2$ near $\lambda$$=$$0$ can be calculated
using the variational properties of $\overline{n}(z,\lambda)$ in CPA and the
Lloyd formula.\cite{Ducastelle,Bruno}

\section{Non-Local CPA for the Double Exchange Model}
\label{sec:cluster_CPA}

  In this section we discuss some general aspects of the non-local CPA to
the problem of orientational spin disorder in the DE model specified by
the Hamiltonian (\ref{eqn:DE_Hamiltonian}). We attempt to
describe the disordered system in an average
by introducing an effective energy-dependent
Hamiltonian,
\begin{equation}
\overline{\cal H}_{\bf ij}(z) = \Sigma_{\bf ii}(z)\delta_{\bf ij} -
\Sigma_{\bf ij}(z)(1-\delta_{\bf ij}),
\label{eqn:disorder_H}
\end{equation}
where $\Sigma_{\bf ij}$ is the non-local part of the self energy, which is
restricted by the nearest neighbors; and $\Sigma_{\bf ii}$ is the local
(site-diagonal) part. The non-local formulation of CPA is
essential because in the low-temperature limit
$\overline{\cal H}_{\bf ij}$ should be replaced by the conventional
kinetic term in which $\Sigma_{\bf ij}$ plays a role of the bare transfer
integral $t_{\bf ij}$.
Therefore, $\Sigma_{\bf ij}$ cannot be
omitted. On the other hand, $\Sigma_{\bf ii}$ will be needed
in order to formulate a
closed system of CPA equations.

\begin{figure}
\centering
\noindent
\epsfxsize=6.5cm  \epsfbox{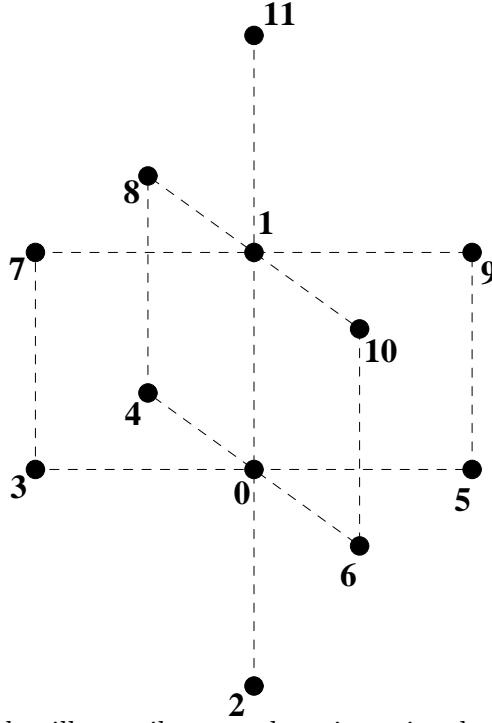}
\caption{Atomic sites which will contribute
         to the orientational average of the $T$-matrix for
         the dimer ${\bf 0}$-${\bf 1}$, when $T(z)$ is given by the
         second-order perturbation theory expression -- Eq.~(\protect\ref{eqn:Tsecond}).}
\label{fig.cluster}
\end{figure}

  We require the effective Hamiltonian (\ref{eqn:disorder_H})
to preserve the cubic symmetry of the system and be translationally
invariant. The first requirement has a different form depending on the
degeneracy of the problem and the symmetry properties of
basis orbitals,
and will be considered separately for the one-orbital
and degenerate DE models.
In any case, using the symmetry properties, all matrix elements of the
self energy $\{ \Sigma_{\bf ii}, \Sigma_{\bf ij} \}$ on the cubic lattice
can be expressed through $\{ \Sigma_{\bf 00}, \Sigma_{\bf 01} \}$
for one of the dimers (for example, ${\bf 0}$-${\bf 1}$
in Fig.~\ref{fig.cluster}).
Then, the Hamiltonian (\ref{eqn:disorder_H})
can be Fourier transformed to the reciprocal space,
$\overline{\cal H}_{\bf q}(z)$$=$$\sum_{\bf j}e^{-i{\bf q} \cdot ({\bf R_i}-{\bf R_j})}
\overline{\cal H}_{\bf ij}(z)$, and
the first equation for the orientationally
averaged Green function can be written as
\begin{equation}
\overline{G}_{\bf ij}(z) = \frac{1}{\Omega_{\rm BZ}} \int d{\bf q}
e^{i{\bf q} \cdot ({\bf R_i}-{\bf R_j})}\left[z-\overline{\cal H}_{\bf q}(z)\right]^{-1},
\label{eqn:disorder_G}
\end{equation}
where the integration goes over the first Brillouin zone with the volume $\Omega_{\rm BZ}$.

  Unfortunately, the non-local form of the Hamiltonian (\ref{eqn:disorder_H}) in the
combination with the translational invariance do not necessarily guaranty the
fulfillment of causality principle, that is the single-particle Green function
should be analytic in the upper half of complex energy plane and satisfy a certain
number of physical requirements.\cite{NickelButler,Elliott,HamadaMiwa,Hettler,Kotliar}
This is still a largely unresolved problem, despite numerous efforts over
decades. We do not have a general solution to it either.
What we try to do here is simply to investigate the behavior of this particular
model and try to answer the question whether it is physical or not.
Note also that neither dynamical cluster approximation\cite{Hettler}
nor cellular DMFT method\cite{Kotliar},
for which the causality can be rigorously proven, can be easily applied
to the problem of off-diagonal disorder.

  The calculations near Van Hove singularities in the case of degenerate DE model
requires very accurate integration in the reciprocal space. In the present work we used
the mesh consisting of 374660 nonequivalent ${\bf q}$-points and corresponding
to $258$$\times$$258$$\times$$258$ divisions of the reciprocal lattice vectors.

  In order to formulate the CPA equations we consider only site-diagonal and
nearest-neighbor elements of $\overline{G}_{\bf ij}$. Again, using the symmetry
properties they can be expressed through
$\{ \overline{G}_{\bf 00}, \overline{G}_{\bf 01} \}$.
In addition, there is a simple relation between $\overline{G}_{\bf 00}$ and
$\overline{G}_{\bf 01}$ for given $\Sigma_{\bf 00}$ and $\Sigma_{\bf 01}$:
\begin{equation}
\overline{G}_{\bf 00}(z)\left[z-\Sigma_{\bf 00}(z)\right] + \sum_{\bf i}
\overline{G}_{\bf 0i}(z)\Sigma_{\bf i0}(z)=1,
\label{eqn:identity}
\end{equation}
which follows from the definition of the Green function (\ref{eqn:disorder_G})
and the Hamiltonian (\ref{eqn:disorder_H}).\cite{PRB01}
Using
this identity, some matrix elements of the Green
function can be easily excluded from
CPA equations.

  In order to obtain the closed system of CPA equations
which connects $\{ \Sigma_{\bf 00}, \Sigma_{\bf 01} \}$
with $\{ \overline{G}_{\bf 00}, \overline{G}_{\bf 01} \}$
we construct the
$T$-matrix:\cite{Ducastelle}
\begin{equation}
\widehat{T}(z)=\left[\widehat{\cal H}-\widehat{\overline{\cal H}}(z)\right]
\left\{\widehat{1}-\widehat{\overline{G}}(z)
\left[\widehat{\cal H}-\widehat{\overline{\cal H}}(z)\right] \right\}^{-1},
\label{eqn:Tgeneral}
\end{equation}
and require the average of scattering due to the fluctuations
$\Delta \widehat{\cal H}$$=$$\widehat{\cal H}$$-$$\widehat{\overline{\cal H}}$
to vanish on every site and every bond of the system, i.e.:\cite{Oguchi,Ducastelle}
\begin{equation}
\overline{T}_{\bf 00}(z)=\overline{T}_{\bf 01}(z)=0.
\label{eqn:CPAconditions}
\end{equation}
The hat-symbols in Eq.~(\ref{eqn:Tgeneral}) means that all the quantities are
infinite matrices in the real space and the matrix multiplications imply also
summation over the intermediate sites.
Note that our approach is different from the so-called
cluster-CPA\cite{NickelButler,Elliott,HamadaMiwa,LichtKats}
because the matrix operations in Eq.~(\ref{eqn:Tgeneral}) are not
confined within a
finite cluster (the dimer, in our case). We believe that our approach is more
logical and more consistent with the requirements of the cubic symmetry
and the translational invariance of the system, because all dimers are
equivalent and should equally contribute to the averaged $T$-matrix. This equivalence is
artificially broken in the cluster-CPA approach, which takes into account the
contributions of only those atoms which are confined within the cluster.
However, our approach also causes some additional
difficulties,
because the non-local fluctuations
$\Delta \widehat{\cal H}$ tend to couple an infinite number of sites in
Eq.~(\ref{eqn:Tgeneral}).
Therefore, for the practical purposes
we restrict ourselves by the perturbation theory expansion
up to the second order of $\Delta \widehat{\cal H}$:
\begin{equation}
\widehat{T}(z) \simeq \left[\widehat{\cal H}-\widehat{\overline{\cal H}}(z)\right] +
\left[\widehat{\cal H}-\widehat{\overline{\cal H}}(z)\right] \widehat{\overline{G}}(z)
\left[\widehat{\cal H}-\widehat{\overline{\cal H}}(z)\right].
\label{eqn:Tsecond}
\end{equation}
As we will show, the first term in this expansion corresponds to the approximation
considered by de Gennes,\cite{deGennes} and the next term is the
first correction to this approximation.
In order to evaluate the matrix elements $\overline{T}_{\bf 00}(z)$ and
$\overline{T}_{\bf 01}(z)$ in the approximation given by Eq.~(\ref{eqn:Tsecond})
it is necessary to consider the interactions confined within the
twelve-atom
cluster shown in Fig.~\ref{fig.cluster}
(obviously, an additional term in the
perturbation theory
expansion for the $T$-matrix would
require a bigger cluster). All such contributions are listed in
Table~\ref{tab:terms}.

\section{One-orbital Double Exchange Model}
\label{sec:1band}

  In the one-orbital case,
the effective DE Hamiltonian
takes the following form, in the reciprocal space:
$$
\overline{\cal H}_{\bf q}(z)=\Sigma_{\bf 00}(z) - 2
(c_x + c_y + c_z) \Sigma_{\bf 01}(z),
$$
where $\Sigma_{\bf 00}(z)$ and $\Sigma_{\bf 01}(z)$ are $C$-numbers,
$c_\gamma$$=$$\cos q_\gamma$, and all energies throughout in this section
are in units of the effective transfer integral $t_0$, which is related with
the $e_g$-bandwidth $W$ in the FM state as $t_0$$=$$W/12$.
Elements of the Green function, $\overline{G}_{\bf 00}(z)$ and
$\overline{G}_{\bf 01}(z)$, are obtained from Eq.~(\ref{eqn:disorder_G}).
In subsequent derivations we will retain both
$\overline{G}_{\bf 00}(z)$ and
$\overline{G}_{\bf 01}(z)$, but only for the sake of convenience of the notations, because
formally $\overline{G}_{\bf 01}(z)$ can be expressed through $\overline{G}_{\bf 00}(z)$
using identity (\ref{eqn:identity}).

\begin{figure}
\centering
\noindent
\epsfxsize=10cm  \epsfbox{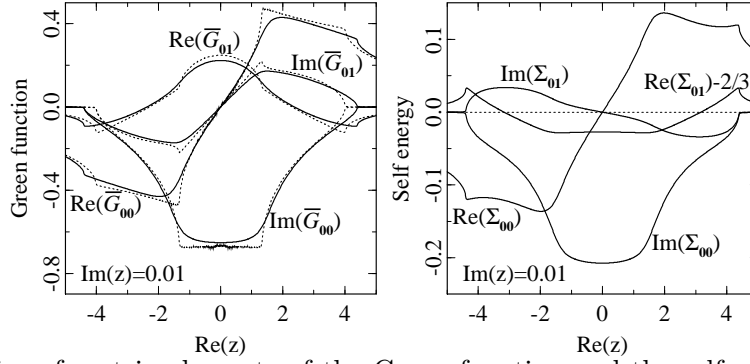}
\caption{Behavior of matrix elements of the Green function and the self energy
         along the real axis
         in the one-orbital double exchange model
         (in units of transfer integral $t_0$).
         The same
         results, but using
         the first-order expression for the $T$-matrix
         with respect to the fluctuations
         $\Delta{\cal H}$
         are shown by
         dotted line.}
\label{fig.1bresults}
\end{figure}

  The self-consistent CPA equations are obtained from the conditions
$\overline{T}_{\bf 00}(z)$$=$$\overline{T}_{\bf 01}(z)$$=$$0$.
In the local coordinate frame associated with the
site ${\bf 0}$, ${\bf e}'_{\bf 0}$$=$$(0,0,1)$
and
all contributions to $T_{\bf 00}(z)$ and $T_{\bf 01}(z)$
shown in Table~\ref{tab:terms} should be averaged
over
the directions of magnetic moments
of remaining sites of the cluster with probability functions
given by Eq.~(\ref{eqn:dfunction}).
This is a tedious, but rather straightforward procedure.
In order to calculate the thermal averages, it is useful to remember
identities listed in Ref.~\onlinecite{comment.1}.
We drop here all
details and, just for reader's convenience, list in Appendix~\ref{apndx:1}
the averaged values for all contributions shown in Table~\ref{tab:terms}
without the derivation, so that every step can be easily checked.
We also introduce short notations for the self energies:
$\Sigma_{\bf 00}$$\equiv$$\sigma_0$ and
$\Sigma_{\bf 01}$$\equiv$$2/3$$+$$\sigma_1$,
and for the Green function: $\overline{G}_{\bf 00}$$\equiv$$g_0$ and
$\overline{G}_{\bf 01}$$\equiv$$g_1$.
Then, the CPA equations $\overline{T}_{\bf 00}(z)$$=$$0$ and
$\overline{T}_{\bf 01}(z)$$=$$0$ can be written in the form
(for $\alpha$$=$$0$ and $1$, respectively):
\begin{equation}
\sigma_\alpha=\Phi_\alpha(\sigma,g)+\Psi_\alpha(\sigma,g)\lambda^2,
\label{eqn:CPAequations}
\end{equation}
where
$$
\Phi_0(\sigma,g)=\left(\sigma_0^2+6\sigma_1^2+\frac{1}{3}\right)g_0
-12\sigma_0\sigma_1g_1,
$$
$$
\Phi_1(\sigma,g)=2\sigma_0\sigma_1g_0-\left(\sigma_0^2+15\sigma_1^2
+2\sigma_1+\frac{13}{54}\right)g_1,
$$
$$
\Psi_0(\sigma,g)=\frac{8}{15}\left(\frac{}{} \sigma_0g_1-\sigma_1g_0\right)
-\frac{1}{45}g_0,
$$
and
$$
\Psi_1(\sigma,g)=\frac{2}{45}-\frac{4}{45}\sigma_0g_0+
\left(\frac{22}{15}\sigma_1-\frac{7}{48}\right)g_1.
$$
These equations should be solved self-consistently in combination
with the definition (\ref{eqn:disorder_G}) for the Green function.
Different elements of $\overline{G}_{\bf ij}(z)$ and $\Sigma_{\bf ij}(z)$
obtained in such a manner
for the PM state ($\lambda$$=$$0$) are shown
in Fig.~\ref{fig.1bresults}.
We note the following:
\begin{enumerate}
\item
both $\overline{G}_{\bf ij}(z)$ and $\Sigma_{\bf ij}(z)$ are analytic in the
upper half of the complex plane;
\item
${\rm Im}\overline{G}_{\bf 00}(z)$$\leq$$0$ in the upper half plane;
\item
the (numerically obtained) integrated density of states lies in the interval
$0$$\leq$$\overline{n}(\mu)$$\leq$$1$ and takes all values within this
interval as the function of chemical potential $\mu$, meaning that our
system is well defined for all physical values of the electronic density.
\end{enumerate}
We believe that for our purposes, the fulfillment of these three causality principles
is quite sufficient and some additional requirements
do not necessary apply here.\cite{comment.7}
Note also that in the one-orbital case there is only one CPA solution
in the complex energy plane.

  In first order expansion for the $T$-matrix
with respect to $\Delta {\cal H}$
(thereafter all quantities corresponding to such
approximation will be denoted by tilde), we naturally reproduce
parameters of the DE model by Gennes:\cite{deGennes}
$\tilde{\Phi}_0$$=$$\tilde{\Phi}_1$$=$$\tilde{\Psi}_0$$=$$0$ and
$\tilde{\Psi}_1$$=$$2/45$.
The corresponding elements of the Green function are also
shown in Fig.~\ref{fig.1bresults}.
Although the second-order approach gives rise to the
large matrix elements of the self energy,
the elements of the Green function obtained in the
first and second order with respect to $\Delta {\cal H}$ are
surprisingly close (apart from a broadening in the second-order
approach, caused by the imaginary part of the self energy),
meaning that there is a good deal of cancellations of
different contributions to $\overline{G}_{\bf 00}$ and $\overline{G}_{\bf 01}$.

  However, it is not true for the Curie temperatures.
While in the first order $T_C$ is solely determined by $\overline{G}_{\bf 00}$,
in the second-order it explicitly depends on both $\overline{G}_{\bf 00}$
and matrix
elements of the self energy, which make significant
difference from the canonical behavior.\cite{deGennes}

  Indeed, $T_C$ can be obtained from Eq.~(\ref{eqn:TC}).
In order to evaluate the DE energy,
we start with the PM solution ($\lambda$$=$$0$)
and include all contributions of the first order
of $\lambda^2$ as a perturbation.
Employing variational properties of the averaged integrated density of
states,\cite{Ducastelle,Bruno}
$\overline{n}(z,\lambda)$ can be found using the Lloyd formula:
\begin{equation}
\overline{n}(z,\lambda) \simeq \overline{n}(z,0) + \frac{1}{\pi}
{\rm Im}
\left\{ \Psi_0(z)g_0(z)-6  \Psi_1(z)g_1(z) \right\} \lambda^2,
\label{eqn:Lloyd}
\end{equation}
where
$\Psi_0$, $\Psi_1$, $g_0$, and $g_1$ correspond to the
PM state.\cite{comment.8}
Moreover, $g_1(z)$ can be expressed through $g_0(z)$
using identity (\ref{eqn:identity}).
Then, the DE energy takes the form:
$E_D(T,\lambda)$$\simeq$$E_D(T,0)$$+$${\cal D}(T)\lambda^2$, where
\begin{equation}
{\cal D}(T)=-\frac{1}{\pi} {\rm Im} \int_{-\infty}^{+\infty} dz
f_T(z-\mu) \left\{ \left( \Psi_0(z) + \frac{\Psi_1(z)[z-\sigma_0(z)]}
{\sigma_1(z)+2/3}\right)g_0(z) - \frac{\Psi_1(z)}{\sigma_1(z)+2/3} \right\}.
\label{eqn:DvT}
\end{equation}
Taking into account the explicit expression for the entropy
term, Eq.~(\ref{eqn:entropy}), we arrive at the following
equation:
\begin{equation}
k_B T_C = -6 {\cal D}(T_C).
\label{eqn:eqTC}
\end{equation}

  In the first order with respect to $\Delta {\cal H}$
(corresponding to the choice
$\tilde{\sigma}_0$$=$$\tilde{\sigma}_1$$=$$\tilde{\Psi}_0$$=$$0$ and
$\tilde{\Psi}_1$$=$$2/45$) and after replacing
$f_T(z$$-$$\mu)$ by $\Theta(z$$-$$\mu)$,
$\tilde{\cal D}(T)$
can be expressed through the DE energy of the PM state:
$\tilde{\cal D}$$=$$-$$\frac{1}{15}\tilde{E}_D({\rm PM})$.
Taking into account that
$\tilde{E}_D({\rm PM})$$=$$\frac{2}{3}E_D({\rm FM})$,
we arrive at the well
known expression obtained by de Gennes:\cite{deGennes}
$k_B \tilde{T}_C$$=$$-$$\frac{4}{15}E_D({\rm FM})$,
where $E_D({\rm FM})$ is the DE energy of the fully polarized FM state.

\begin{figure}
\centering
\noindent
\epsfxsize=10cm  \epsfbox{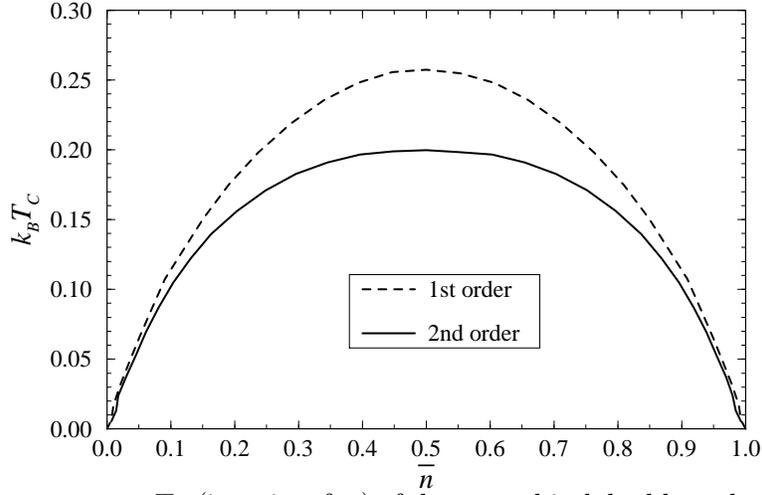}
\caption{Curie temperature $T_C$
         (in units of $t_0$)
         of the one-orbital
         double exchange model obtained using the first- and second-order
         expression for the $T$-matrix with respect to the
         fluctuations $\Delta{\cal H}$
         as a function of averaged electronic density.}
\label{fig.1bTC}
\end{figure}

  The results of calculations are shown in Fig.~\ref{fig.1bTC}.\cite{comment.9}
A more accurate treatment of $\Delta {\cal H}$ in the
expression for the $T$-matrix significantly reduces $T_C$ (up to 20\% in the
second-order approach).
The values $T_C$$\simeq$$0.20t_0$ and $0.17t_0$ obtained correspondingly at
$\overline{n}$$=$$0.5$ and $\overline{n}$$=$$0.25$
are substantially reduced in comparison with
the (local) DMFT approach.\cite{comment.10}
We also note a good agreement with the results by
Alonso {\it et al.},\cite{Alonso} who used similar
variational mean-filed approach
supplemented with
the moments-method for computing the averaged density of states
in the one-orbital DE model.
Using the value
$W$$\approx$$4$eV obtained in
bands structure calculations for the FM state,\cite{Springer02}
$T_C$ can be roughly estimated as
$T_C$$\leq$$800$K. The upper bound,
corresponding to $\overline{n}$$=$$0.5$,
exceeds the experimental data by factor two.
$T_C$ can be further reduced by taking into consideration
the antiferromagnetic (AFM) superexchange (SE) interactions between the localized
spins\cite{deGennes,Alonso,Dzero,comment.3}
and spatial spin correlations.
For example, according to recent Monte Carlo simulations,
the latter can reduce $T_C$ up to $0.12t_0$$\approx$$460$K at
$\overline{n}$$=$$0.5$.\cite{Motome}

\section{Degenerate Double Exchange Model for the \lowercase{$e_g$} Electrons}
\label{sec:2band}

  In this section we consider a more realistic example of the double exchange model
involving two $e_g$ orbitals, which have the following order:
$|1\rangle$$\equiv$$|x^2$$-$$y^2\rangle$ and $|2\rangle$$\equiv$$|3z^2$$-$$r^2\rangle$.
Then all quantities,
such as the transfer integrals ${\bf t_{ij}}$, the averaged Green function
$\overline{\bf G}_{\bf ij}(z)$, and the self energy $\mbox{\boldmath$\Sigma$}_{\bf ij}(z)$
become $2$$\times$$2$ matrices in the basis of these orbitals
(thereafter, the bold symbols will be reserved for such matrix notations).
The DE model is
formulated in the same way as in the one-orbital case. Namely, modulations of the
transfer integrals are described by Eq.~(\ref{eqn:DE_Hamiltonian}) with the complex
multipliers $\xi_{\bf ij}$. What important is the peculiar form of the
${\bf t_{ij}}$ matrices
on the cubic lattice, given in terms of Slater-Koster
integrals of the $dd\sigma$ type.\cite{SlaterKoster}
For example, for the ${\bf 0}$-${\bf 1}$ and ${\bf 0}$-${\bf 2}$ bonds
parallel to the $z$-axis (see Fig.~\ref{fig.cluster}) these matrices have the form:
\begin{equation}
{\bf t_{01}}={\bf t_{02}} = \left(
\begin{array}{cc}
0 & 0 \\
0 & 1 \\
\end{array}
\right).
\label{eqn:t01}
\end{equation}
In the other words, the hoppings along the $z$-direction are allowed only between
the $3z^2$$-$$r^2$ orbitals.
Throughout in this section, the absolute values of the parameter
$(dd\sigma )$$=$$W/6$$\approx$$0.7$eV
will be used as the energy unit.

  The remaining matrix elements in the $xy$-plane
can be obtained by 90$^\circ$ rotations of the ${\bf 0}$-${\bf 1}$ bond
around the
$x$- and $y$-axes. Corresponding transformations of the $e_g$ orbitals are
given by:
\begin{equation}
\mbox{\boldmath${\cal U}$}_x = \frac{1}{2} \left(
\begin{array}{cc}
     1    & -\sqrt{3} \\
-\sqrt{3} &     -1    \\
\end{array}
\right),
\label{eqn:ux}
\end{equation}
and
\begin{equation}
\mbox{\boldmath${\cal U}$}_y = \frac{1}{2} \left(
\begin{array}{cc}
     1    &  \sqrt{3} \\
 \sqrt{3} &     -1    \\
\end{array}
\right),
\label{eqn:uy}
\end{equation}
respectively.
Then, it is easy to obtain the well known expressions:\cite{SlaterKoster}
\begin{equation}
{\bf t_{04}}={\bf t_{06}}= \mbox{\boldmath${\cal U}$}_y^T
{\bf t_{01}}\mbox{\boldmath${\cal U}$}_y =
\frac{1}{4}
\left(
\begin{array}{cc}
     3    & -\sqrt{3} \\
-\sqrt{3} &      1    \\
\end{array}
\right),
\label{eqn:t04}
\end{equation}
for the hoppings parallel to the $x$-axis, and
\begin{equation}
{\bf t_{03}}={\bf t_{05}}= \mbox{\boldmath${\cal U}$}_x^T
{\bf t_{01}}\mbox{\boldmath${\cal U}$}_x =
\frac{1}{4}
\left(
\begin{array}{cc}
     3    &  \sqrt{3} \\
 \sqrt{3} &      1    \\
\end{array}
\right),
\label{eqn:t03}
\end{equation}
for the hoppings parallel to the $y$-axis.

  We remind these symmetry properties
as an introduction to the analysis of the self energy
in the case of the spin disorder,
which
obeys very similar symmetry constraints.
Namely, since the cubic symmetry is not destroyed by
the disorder, the $x^2$$-$$y^2$ and $3z^2$$-$$r^2$ orbitals
belong to the same representation of the point symmetry group.
Therefore,
the site-diagonal part of the self energy (as well as of the Green function)
will be both diagonal and degenerate with respect to the orbital indices, i.e.:
$$
\mbox{\boldmath$\Sigma_{\bf 00}$}=
\left(
\begin{array}{cc}
\Sigma_{\bf 00}^{11} &    0                     \\
         0           &    \Sigma_{\bf 00}^{22}  \\
\end{array}
\right),
$$
where $\Sigma_{\bf 00}^{11}$$=$$\Sigma_{\bf 00}^{22}$.
On the other hand, the matrix elements
associated with the bond ${\bf 0}$-${\bf 1}$
should
transform to themselves according to
the tetragonal ($C_{4v}$) symmetry.
Since the $x^2$$-$$y^2$ and $3z^2$$-$$r^2$ orbitals belong to different
representations of the $C_{4v}$ group ($a_{1g}$ and $b_{2g}$, respectively),
the corresponding non-local part of the self energy has the form:
$$
\mbox{\boldmath$\Sigma_{\bf 01}$}=
\mbox{\boldmath$\Sigma_{\bf 02}$}=
\left(
\begin{array}{cc}
\Sigma_{\bf 01}^{11} &    0                     \\
         0           &    \Sigma_{\bf 01}^{22}  \\
\end{array}
\right).
$$
Note that
$\Sigma_{\bf 01}^{11}(z)$ is not necessarily zero.
The identity $t_{\bf 01}^{11}$$=$$0$, which holds for the
transfer integrals, reflects the hidden symmetry of the ordered
FM state. However, there is no reason to expect that
the same identity
will be preserved in the case of the spin disorder.
Moreover,
as we will show below, the condition $\Sigma_{\bf 01}^{11}(z)$$\neq$$0$ is
indispensable in order to formulate the closed system of CPA equations.

  Thus, in the case of the orbital degeneracy,
there are {\it three} independent matrix
elements of the self energy: $\Sigma_{\bf 00}^{11}(z)$$=$$\Sigma_{\bf 00}^{22}(z)$,
$\Sigma_{\bf 01}^{11}(z)$ and $\Sigma_{\bf 01}^{22}(z)$.
In comparison with the one-orbital case we gain an additional
non-local parameter, which may control the properties of disordered
DE
systems, even in the case of the strictly imposed cubic symmetry.

  The matrix elements of the self energy in the $xy$-plane can be obtained
using the transformations (\ref{eqn:ux}) and (\ref{eqn:uy}), which yield:
$$
\mbox{\boldmath$\Sigma_{\bf 04}$}=
\mbox{\boldmath$\Sigma_{\bf 06}$}=
\mbox{\boldmath${\cal U}$}_y^T
\mbox{\boldmath$\Sigma_{\bf 01}$}
\mbox{\boldmath${\cal U}$}_y =
\frac{1}{4}
\left(
\begin{array}{cc}
 \Sigma_{\bf 01}^{11} +3\Sigma_{\bf 01}^{22}         & -\sqrt{3}(\Sigma_{\bf 01}^{22}-\Sigma_{\bf 01}^{11}) \\
-\sqrt{3}(\Sigma_{\bf 01}^{22}-\Sigma_{\bf 01}^{11}) &   3\Sigma_{\bf 01}^{11} +\Sigma_{\bf 01}^{22}  \\
\end{array}
\right)
$$
and
$$
\mbox{\boldmath$\Sigma_{\bf 03}$}=
\mbox{\boldmath$\Sigma_{\bf 05}$}=
\mbox{\boldmath${\cal U}$}_x^T
\mbox{\boldmath$\Sigma_{\bf 01}$}
\mbox{\boldmath${\cal U}$}_x =
\frac{1}{4}
\left(
\begin{array}{cc}
 \Sigma_{\bf 01}^{11} +3\Sigma_{\bf 01}^{22}         &  \sqrt{3}(\Sigma_{\bf 01}^{22}-\Sigma_{\bf 01}^{11}) \\
 \sqrt{3}(\Sigma_{\bf 01}^{22}-\Sigma_{\bf 01}^{11}) &   3\Sigma_{\bf 01}^{11} +\Sigma_{\bf 01}^{22}  \\
\end{array}
\right).
$$
Then, the effective Hamiltonian takes the following
form, in the reciprocal space:
$$
\overline{\mbox{\boldmath${\cal H}$}}_{\bf q} =
\mbox{\boldmath$\Sigma_{\bf 00}$} - \frac{1}{2}
\left(
\begin{array}{cc}
(\Sigma_{\bf 01}^{11} +3\Sigma_{\bf 01}^{22})(c_x + c_y) +4\Sigma_{\bf 01}^{11}
c_z &
\sqrt{3}(\Sigma_{\bf 01}^{22}-\Sigma_{\bf 01}^{11})(c_y - c_x) \\
\sqrt{3}(\Sigma_{\bf 01}^{22}-\Sigma_{\bf 01}^{11})(c_y - c_x) &
(3\Sigma_{\bf 01}^{11} +\Sigma_{\bf 01}^{22})(c_x + c_y) +4\Sigma_{\bf 01}^{22}
c_z \\
\end{array}
\right).
$$
The orientationally averaged Green function can be
then obtained from Eq.~(\ref{eqn:disorder_G}).
Similar to the self energy, all matrix elements of the Green function
can be expressed through $\overline{G}_{\bf 00}^{11}$, $\overline{G}_{\bf 01}^{11}$,
and $\overline{G}_{\bf 01}^{22}$, using the symmetry properties (\ref{eqn:t04}) and (\ref{eqn:t03}).
In addition, they satisfy the matrix equation (\ref{eqn:identity}), from which one can easily
exclude one of the matrix elements.

  The self-consistent CPA equations are obtained from the conditions
(\ref{eqn:CPAconditions}). We drop here all details
and present only the final result (some intermediate
expressions for thermal averages of different contributions listed in
Table~\ref{tab:terms} are given in Appendix~\ref{apndx:2}).
We also introduce short notations for the self-energy:
$\Sigma_{\bf 00}^{11}$$=$$\Sigma_{\bf 00}^{22}$$\equiv$$\sigma_0$,
$\Sigma_{\bf 01}^{11}$$\equiv$$\sigma_1$,
and $\Sigma_{\bf 01}^{22}$$\equiv$$2/3$$+$$\sigma_2$;
and for the Green functions:
$\overline{G}_{\bf 00}^{11}$$\equiv$$g_0$, $\overline{G}_{\bf 01}^{11}$$\equiv$$g_1$,
and $\overline{G}_{\bf 01}^{22}$$\equiv$$g_2$.
Then, the CPA equations can be presented in the form (\ref{eqn:CPAequations})
with $\alpha$$=$ 0, 1, and 2 corresponding to the conditions
$T_{\bf 00}^{11}(z)$$=$$0$, $T_{\bf 01}^{11}(z)$$=$$0$, and $T_{\bf 01}^{22}(z)$$=$$0$,
respectively.
Using the second-order expression for the $T$-matrix -- Eq.~(\ref{eqn:Tsecond}),
one can obtain the following expressions for
the coefficients $\Phi_\alpha$ and $\Psi_\alpha$:
\begin{equation}
\Phi_0(\sigma,g)=\left(\sigma_0^2+3\sigma_1^2+3\sigma_2^2+\frac{1}{6}\right)g_0
-6\sigma_0\sigma_1g_1-6\sigma_0\sigma_2g_2,
\label{eqn:CPA2.1}
\end{equation}
\begin{eqnarray}
\Phi_1(\sigma,g)= & 2\sigma_0\sigma_1g_0 - &
\left(\sigma_0^2+\frac{21}{4}\sigma_1^2+\frac{9}{4}\sigma_2^2+\frac{3}{2}\sigma_1\sigma_2+
\frac{1}{6}\sigma_1 + \frac{1}{2}\sigma_2\right)g_1 - \nonumber \\
 & &
\left(\frac{3}{4}\sigma_1^2+\frac{3}{4}\sigma_2^2+\frac{9}{2}\sigma_1\sigma_2-
\frac{1}{6}\sigma_1+\frac{1}{6}\sigma_2+\frac{2}{3}\right)g_2,
\label{eqn:CPA2.2}
\end{eqnarray}
\begin{eqnarray}
\Phi_2(\sigma,g)= & 2\sigma_0\sigma_2g_0 - &
\left(\frac{3}{4}\sigma_1^2+\frac{3}{4}\sigma_2^2+\frac{9}{2}\sigma_1\sigma_2-
\frac{1}{6}\sigma_1+\frac{1}{6}\sigma_2 \right)g_1 - \nonumber \\
 & &
\left(\sigma_0^2+\frac{9}{4}\sigma_1^2+\frac{21}{4}\sigma_2^2+\frac{3}{2}\sigma_1\sigma_2+
\frac{1}{6}\sigma_1 + \frac{1}{2}\sigma_2 + \frac{7}{54} \right)g_2,
\label{eqn:CPA2.3}
\end{eqnarray}
\begin{equation}
\Psi_0(\sigma,g)=\frac{4}{15}\left(\frac{}{}\sigma_0g_2-\sigma_2g_0\right)-\frac{1}{90}g_0,
\label{eqn:CPA2.4}
\end{equation}
\begin{equation}
\Psi_1(\sigma,g)= \frac{1}{90} \left\{ \left( \frac{}{} 7\sigma_1+21\sigma_2+2\right)g_1 +
\left(21\sigma_1+7\sigma_2+\frac{2}{3}\right)g_2 \right\},
\label{eqn:CPA2.5}
\end{equation}
and
\begin{equation}
\Psi_2(\sigma,g)=
\frac{2}{45} -\frac{4}{45}\sigma_0g_0 +
\frac{1}{90} \left\{ \left( 17\sigma_1+7\sigma_2+\frac{2}{3}\right)g_1 +
\left(7\sigma_1+45\sigma_2+\frac{7}{3}\right)g_2 \right\}.
\label{eqn:CPA2.6}
\end{equation}

\subsection{CPA solution for the paramagnetic phase}
\label{sec:para}

  In this section we consider CPA solutions for the PM phase
of the degenerate DE model,
and argue that the situation is {\it qualitatively different} from the
one-orbital model, even in the case of cubic symmetry.

\begin{figure}
\centering
\noindent
\epsfxsize=10cm  \epsfbox{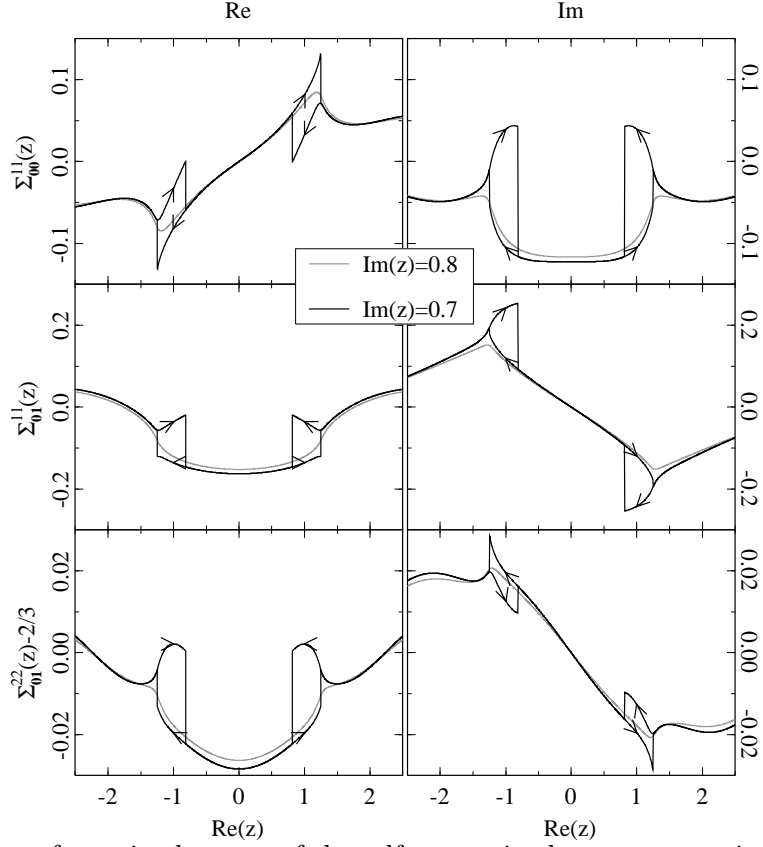}
\caption{Behavior of matrix elements of the self energy in
         the paramagnetic state of degenerate double exchange.
         For ${\rm Im}(z)$$\geq$$0.75$ there is only
         one solution, while for ${\rm Im}(z)$$<$$0.75$ one can obtain
         two self-consistent CPA solutions in certain interval of ${\rm Re}(z)$
         by starting the iterations
         with the self energy obtained for the previous value of ${\rm Re}(z)$
         and moving either in the positive or negative direction of the
         real axis (shown as a hysteresis).}
\label{fig.2bselfe}
\end{figure}

\begin{figure}
\centering
\noindent
\epsfxsize=8cm  \epsfbox{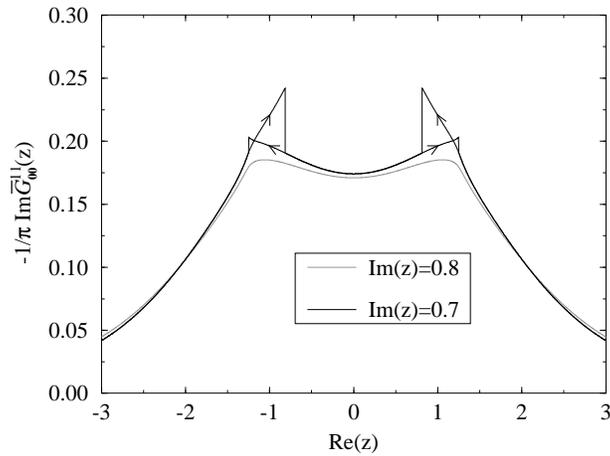}
\caption{Imaginary part of the Green function (the local density of states)
         for the paramagnetic state of
         degenerate double exchange model. See Fig.~\protect\ref{fig.2bselfe}
         for description.}
\label{fig.2bgreen}
\end{figure}

  Let us first consider the limit
$|z|$$\rightarrow$$\infty$. In this case, the matrix elements of the
Green function have the following asymptotic behavior:\cite{Heine}
$g_0(z)$$\rightarrow$$\frac{1}{z}$,
$g_1(z)$$\rightarrow$$\frac{A}{z^2}$, and
$g_2(z)$$\rightarrow$$\frac{B}{z^2}$;
and the CPA solution in the second order of $\frac{1}{z}$ can be easily obtained
analytically from Eqs.~(\ref{eqn:CPA2.1})-(\ref{eqn:CPA2.3}) as
$\sigma_0(z)$$\rightarrow$$\frac{1}{6z}$,
$\sigma_1(z)$$\rightarrow$$-$$\frac{2B}{3z^2}$, and
$\sigma_2(z)$$\rightarrow$$-$$\frac{7B}{54z^2}$.
It is a single-valued solution and
in this sense the situation is similar to
the one-orbital case.
The analysis is supported by numerical calculations for
large but finite $|z|$, and represents a typical behavior
when ${\rm Im}(z)$$\geq$$0.75$ (Fig.~\ref{fig.2bselfe}).
The Van Hove singularities
are smeared due to the large imaginary
part of $z$ and the self energy (Fig.~\ref{fig.2bgreen}).

  However, when we start to approach the real axis,
the situation changes dramatically.
In the first calculations of such type
we fix ${\rm Im}(z)$ and solve CPA equations by moving along the real axis
and
each time
starting with the self-consistent self energy obtained for the previous
value of ${\rm Re}(z)$. Then,
in certain region of the complex plane
we obtain {\it two different solutions},
depending on whether we move in
the positive or negative direction of ${\rm Re}(z)$. A typical hysteresis loop
corresponding to ${\rm Im}(z)$$=$$0.7$ is shown in Figs.~\ref{fig.2bselfe} and \ref{fig.2bgreen}.

  The behavior is related with the existence of two Van Hove singularities at
the $(0,\pi,0)$ and $(\pi,\pi,0)$ points of the Brillouin zone, which are responsible
for some sort of instability in the system.
The singularities become increasingly important when
$z$ approaches the real axis.
The positions of these singularities
depend on matrix elements of
the self energy, and given by
${\rm Re}(\Sigma_{\bf 00}^{11}$$-$$3\Sigma_{\bf 01}^{11}$$+$$\Sigma_{\bf 01}^{22})$ and
${\rm Re}(\Sigma_{\bf 00}^{11}$$+$$\Sigma_{\bf 01}^{11}$$-$$3\Sigma_{\bf 01}^{22})$,
respectively.
Therefore, by choosing different starting points for the
self energy, the singularities can be shifted either
'to the right' or 'to the left' (Fig.~\ref{fig.2bgreen}).
Since the CPA equations are non-linear,
this may
stabilize two different solutions.

\begin{figure}
\centering
\noindent
\epsfxsize=7cm  \epsfbox{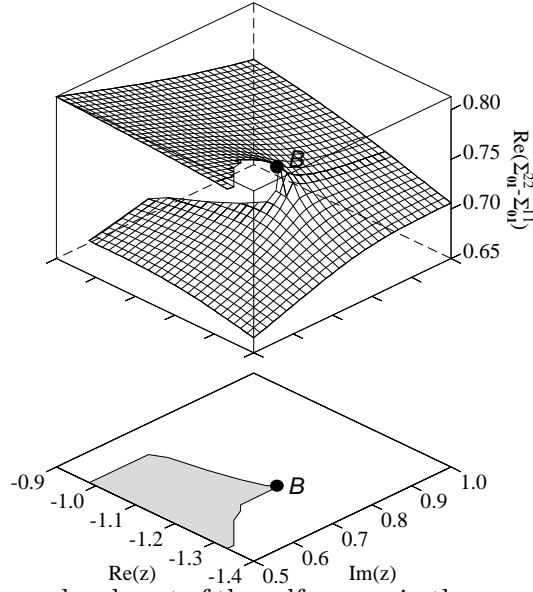}
\caption{Topology of the non-local part of
         the self energy in the complex plane.
         The branch-point is denoted by $B$.
         The projection shows an approximate
         position of the double-valued
         area for the CPA solutions.}
\label{fig.topology}
\end{figure}

  A more complete picture can be
obtained from Fig.~\ref{fig.topology}, where we plot
${\rm Re}(\Sigma_{\bf 01}^{22}$$-$$\Sigma_{\bf 01}^{11})$
in the complex plane.\cite{comment.14}
Depending on the location in the complex plane,
the CPA equations (\ref{eqn:CPAequations}) are again converged to either the same
or two different solutions. The double-valued behavior of
${\rm Re}(\Sigma_{\bf 01}^{22}$$-$$\Sigma_{\bf 01}^{11})$ typically
occurs within the shaded area.
Note that this area is the result of
numerical calculations which depend on the choice of the starting point.
We do not exclude here the possibility that our result may be
incomplete and that with a better choice of the starting conditions the double-valued
area may be enlarged.

  Our analysis is limited by ${\rm Im}(z)$$\approx$$0.5$. Further
attempts to approach the real axis were conjugated with serious difficulties:
the topology of solutions becomes increasingly complicated and may include several additional
branches, some of which are presumably unphysical. At the present stage we do not have a
clear strategy of how to deal with this problem. Of course, formally the situation can
be regarded as the violation of causality principles.\cite{NickelButler,Elliott}
Nevertheless, we would like to believe that
the double-valued behavior of the Green function and the self energy
obtained for ${\rm Im}(z)$$\geq$$0.5$ does have a physical meaning and is not an artifact
of the model analysis.
That is because of the following reasons:
\begin{enumerate}
\item
for both solutions, the Green function satisfies the inequality
${\rm Im}\overline{G}_{\bf 00}^{11}$$\leq$$0$;
\item
the system is well defined in the whole interval of
densities $0$$\leq$$\overline{n}(\mu)$$\leq$$1$ (corresponding to the doping range
$0$$\leq$$x$$\leq$$1$ or the position of the chemical potential $\mu$$\leq$$1$)
{\it if and only if} to take into account the {\it superposition of the two
solutions}. Conversely, by considering only one of the solutions (and
disregarding the other one as unphysical), the density will exhibit
a finite jump (a discontinuity) at certain value of $\mu$, and the system will be
undefined within the discontinuity range.
Presumably, the discontinuity of the electronic density
would
present even more unphysical
behavior
than the fact of the existence of two CPA solutions.
\item
In some sense the appearance of two CPA solutions fits well into the logic of our work.
As it was discussed before,
the non-local part of the self energy
in the degenerate case is characterized by two different matrix elements
and, in comparison with the one-orbital case,
acquires an additional degree of freedom (which may be related
with some non-local order parameter attached to the bond of the system).
At the same time, the positions of Van Hove singularities,
which control the behavior of non-linear CPA equations,
depend on these matrix
elements.
Therefore, the appearance of several CPA solutions seems to be natural.
\end{enumerate}

  Note that ${\rm Im}(z)$$\approx$$0.5$ corresponds to the position
of the first Matsubara pole for $k_BT$$\approx$$0.159$. Taking into account the realistic
value of the parameter $(dd\sigma)$$\simeq$$0.7$eV,\cite{Springer02}
it roughly corresponds to $T$$\approx$$1200$K,
which can be regarded as the lowest estimate for
the temperature for which our analysis is strictly justified.

  Below we discuss possible physical consequences of
the existence
of two CPA solutions in the PM state.
Our scenario
is based on the following observations (see Fig.~\ref{fig.topology}):
\begin{enumerate}
\item
The existence of the branch-point ($B$ in Fig.~\ref{fig.topology}),
which forms two physical branches of CPA solutions in certain area of the
complex plane. The requirement implies that there is a continuous path
around the branch-point, which connects the points located on two
different branches.
\item
Both branches of the (multi-valued) Green function and the self energy
are analytic (perhaps except the branch-point itself and the branch-edges).
The requirement allows us to use the standard theorems of the contour integration:
for example, the contour integral around the branch-point does not depend
on the form of the contour, etc.
\end{enumerate}

  Strictly speaking, the second requirement is a postulate which is
solely based on results of numerical
calculations and at the present stage we do not have a general proof for it.
Assume, it is correct.
Then, the physical interpretation of the multi-valued behavior becomes
rather straightforward
and two CPA solutions can be linked to
two PM phases (with different densities)
corresponding to the same chemical potential $\mu$.
The situation has many things in common with the phenomenon of
inhomogeneous phase
separation, which was intensively discussed for
manganites.\cite{Nagaev,Dagotto}
The new aspect in our case is that {\it both phases are paramagnetic}.
The position of the branch-point itself
can be related with the temperature, below which the PM state becomes
intrinsicly inhomogeneous.

\subsection{energy integration and appearance of two paramagnetic phases}
\label{sec:integral}

\begin{figure}
\centering
\noindent
\epsfxsize=7cm  \epsfbox{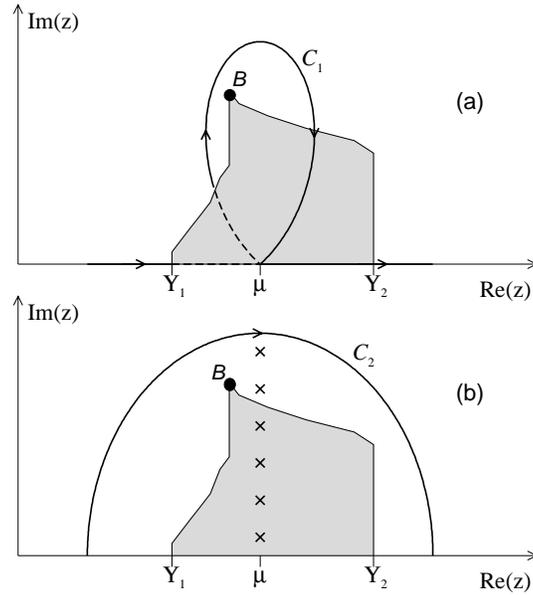}
\caption{Energy integration for the degenerate double exchange model.
         (a): The integral along the real axis plus the discontinuity
         given by the contour integral $C_1$ around the branch-point $B$.
         (b): An equivalent expression in terms of the contour integral $C_2$
         spreading in the single-valued area of the complex plane
         and
         residues calculated at Matsubara poles.
         The latter contributions are different for two different branches,
         that is equivalent to the discontinuity term in the scheme (a).}
\label{fig.cplane}
\end{figure}

  In this section we discuss some aspects of the energy integration
in the complex plane, related
with the existence of two CPA-branches. Let us consider
the integral:
\begin{equation}
X(\mu)=\int_{-\infty}^{+\infty} dz f_T(z-\mu) X(z),
\label{eqn:Tint}
\end{equation}
where $X(\mu)$ is a physical quantity, which can be the density of $e_g$
electrons, the double exchange energy or the change of either of them, and
$X(z)$ has the same topology in the complex plane as the self energy shown
in Fig.~\ref{fig.topology}.
Then, the behavior of the integral (\ref{eqn:Tint}) will depend on the
position of the chemical potential $\mu$ with respect to the double-valued
area. Generally, we should consider three possibilities (see Fig.~\ref{fig.cplane} for
notations):
\begin{enumerate}
\item
$\mu$$<$$\Upsilon_1$. In this case there is only one CPA solution and
the
integral (\ref{eqn:Tint}) can be evaluated by using the standard
techniques
(see, e.g., Ref.~\onlinecite{Wildberger} and references therein).
\item
$\Upsilon_1$$\leq$$\mu$$\leq$$\Upsilon_2$.
In this case the integral (\ref{eqn:Tint}) takes two values for
each value of the
chemical potential $\mu$: $X_1(\mu)$, if the integrand is solely confined within
one branch; and
$X_2(\mu)$$=$$X_1(\mu)$$+$$\Delta X(\mu)$, if it is extended to the second branch.
The discontinuity $\Delta X(\mu)$
is given by the contour integral $C_1$ around the branch-point.
It is important here that we do not try to define $X(z)$ as a single-valued
function by introducing the branch-cuts, which by itself is largely
arbitrary procedure.\cite{NickelButler}
Instead, we treat both branches on an equal footing, which inevitably leads
to the multi-valued behavior of $X(\mu)$.
This integral (\ref{eqn:Tint})
can be replaced by the contour integral $C_2$ spreading in the
single-valued area plus residues calculated at
Matsubara poles $z_n$$=$$\mu$$+$$i\pi k_B T (2n$$+$$1)$:
\begin{equation}
X(\mu)=\int_{C_2} dz f_T(z-\mu)X(z)-2\pi i k_BT \sum_{z_n}X(z_n).
\label{eqn:Tint1}
\end{equation}
In order to calculate $X_1$ and $X_2$, the
Matsubara poles should lie on the first and second
branches of $X(z)$, respectively.\cite{comment.15}
\item
$\mu$$>$$\Upsilon_2$. In this case the integration along the real axis
shall be combined with the
discontinuity $\Delta X$ given by the contour integral $C_1$.
The integration can be replaced by Eq.~(\ref{eqn:Tint1}).
In this case, all Matsubara poles lie on the single branch.
Therefore, $X(\mu)$ is a single-valued function.
\end{enumerate}

\begin{figure}
\centering
\noindent
\epsfxsize=7.5cm  \epsfbox{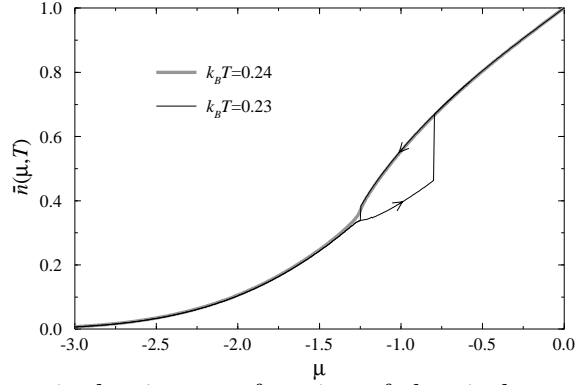}
\caption{Averaged electronic density as a function of chemical potential for the
         paramagnetic state of degenerate double exchange model.
         For $T$$\leq$$0.23$, $\overline{n}(\mu)$ may take two different value
         for the same
         chemical potential $\mu$ which
         correspond to two different CPA solutions
         shown in Fig.~\protect\ref{fig.2bselfe}.}
\label{fig.nef}
\end{figure}

  As an illustration, we show in Fig.~\ref{fig.nef} the behavior of
the
averaged
density as the function of $\mu$.
For $T$$=$$0.24$, corresponding to ${\rm Im}(z_0)$$\simeq$$0.754$,
which is slightly above the branch-point,
the
first Matsubara pole falls beyond the double-valued area
and
$\overline{n}(\mu)$ shows a normal behavior when for each value of
$\mu$ there is only
one value of the electronic density.
For smaller $T$, $\overline{n}(\mu)$ may take two different values
for the same $\mu$. This fact can be interpreted as the
coexistence of two different phases.
Such a behavior typically occurs in the interval
$-$$1.4$$\leq$$\mu$$\leq$$-$$0.8$, which may also depend on the temperature.
As it was mentioned before, any single phase fails to define
the electronic density in the
whole interval $0$$\leq$$\overline{n}(\mu)$$\leq$$1$ because of the
discontinuity of $\overline{n}(\mu)$. For example, for $T$$=$$0.23$
one of the phases is not defined in the interval
$0.34$$\leq$$\overline{n}$$\leq$$0.38$, corresponding to the discontinuity
at $\mu$$=$$-$$1.25$, and the other one - for
$0.46$$\leq$$\overline{n}$$\leq$$0.66$ corresponding to $\mu$$=$$-$$0.80$.
The problem can be resolved only by considering the combination of these
two phases for which $\overline{n}$ is defined everywhere in the
interval $0$$\leq$$\overline{n}(\mu)$$\leq$$1$.

  $T_P$$\approx$$0.23$ (about $1800$K),
roughly corresponding to the position of the
branch-point, can be regarded
as the transition temperature to the two-phase state
(a point of PM phase separation).

\subsection{phase coexistence}
\label{sec:pseparation}

   In this section we
briefly consider the problem of phase coexistence using a
semi-quantitative theory of
non-interactive pseudoalloy.
Namely, we assume that the free energy of the mixed paramagnetic state
is given by
\begin{equation}
{\cal F}_{\rm mix}(y)=(1-y)E_D^{(1)} + y E_D^{(2)} -TS_{\rm mix}(y),
\label{eqn:PS_freee}
\end{equation}
where $E_D^{(1)}$ and $E_D^{(2)}$ are the DE energies of two PM phases
(with lower and higher densities of the
$e_g$ electrons, respectively);
$y$ is the ''alloy concentration''; and $S_{\rm mix}(y)$ is the
configurational mixing entropy:
$$
-TS_{\rm mix}(y) = k_BT\left[y \ln y + (1-y) \ln (1-y) \right].
$$
Then, the equilibrium concentration, which minimizes the free energy
(\ref{eqn:PS_freee}) is given by
$y$$=$$[\exp (\frac{\Delta E_D}{k_BT})$$+$$1]^{-1}$.
The difference of the DE energies $\Delta E_D$$=$$E_D^{(2)}$$-$$E_D^{(1)}$ can be
calculated using the definition (\ref{eqn:DEenergy}) and the formula
(\ref{eqn:Tint1}) for the contour integration. Since the contour $C_2$ is
confined within the single-valued area, $\Delta E_D$ is given by the difference of
residues at a limited number of Matsubara poles:
\begin{equation}
\Delta E_D = 2\pi i k_B T \sum_{z_n} \left[ \overline{n}^{(2)}(z_n)-\overline{n}^{(1)}(z_n) \right].
\label{eqn:DeltaED}
\end{equation}
Some details of these calculations can be found in Appendix~\ref{apndx:3}.
The discontinuity of the electronic density
$\Delta\overline{n}$$=$$\overline{n}^{(2)}$$-$$\overline{n}^{(1)}$ can be
calculated in a similar way.

\begin{figure}
\centering
\noindent
\epsfxsize=7.5cm  \epsfbox{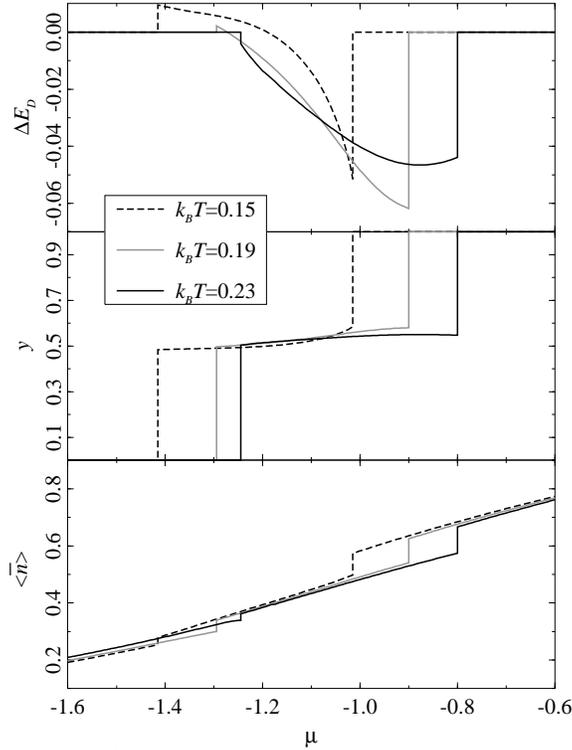}
\caption{Pseudo-alloy picture for the
         two-phase paramagnetic state of the degenerate double exchange model:
         the change of the double exchange energy
         $\Delta E_D$$=$$E^{(2)}_D$$-$$E^{(1)}_D$, the equilibrium
         alloy concentration $y$, and the density of $e_g$ electrons
         averaged simultaneously over the spin orientations and the alloy concentrations
         $ \langle \overline{n} \rangle$$=$$(1$$-$$y)\overline{n}^{(1)}$$+$$y\overline{n}^{(2)}$
         as a function of chemical potential $\mu$ for three
         different
         temperatures (in units of $dd\sigma$-integral).
         The superscripts
         $^{(1)}$ and $^{(2)}$ stand for the phases correspondingly with lower and higher
         electronic densities.}
\label{fig.twophase}
\end{figure}

  Results of these calculations are shown in Fig.~\ref{fig.twophase}.
Particularly, the equilibrium alloy concentration $y$ is close to
0.5, meaning that the difference $\Delta E_D$ is small and the main contribution
to the free energy comes from the entropy term. The electronic density
$\langle \overline{n} \rangle$$=$$(1$$-$$y)\overline{n}^{(1)}$$+$$y\overline{n}^{(2)}$
averaged simultaneously over the spin orientations and the alloy concentrations shows a
discontinuity at the edges of the two-phase region, meaning that the system
is not defined.
This behavior is unphysical and caused by the non-interactive approach to the
problem of phase coexistence. The averaged densities range, for which two PM phases
may
coexist is typically 0.3-0.7
and depends on the temperature.

\subsection{Curie temperature}
\label{sec:TC.2band}

  The Curie temperature can be obtained from Eq.~(\ref{eqn:eqTC}).
In the case of degenerate DE model, the function ${\cal D}(T)$
is given by:\cite{comment.2}
\begin{equation}
{\cal D}(T)= -\frac{1}{\pi} {\rm Im} \int_{-\infty}^{+\infty} dz
f_T(z-\mu) \left\{ 2 \Psi_0(z)g_0(z) -6\left[
\Psi_1(z)g_1(z) + \Psi_2(z)g_2(z) \right] \right\}.
\label{eqn:TCfunc}
\end{equation}

\begin{figure}
\centering
\noindent
\epsfxsize=7.5cm  \epsfbox{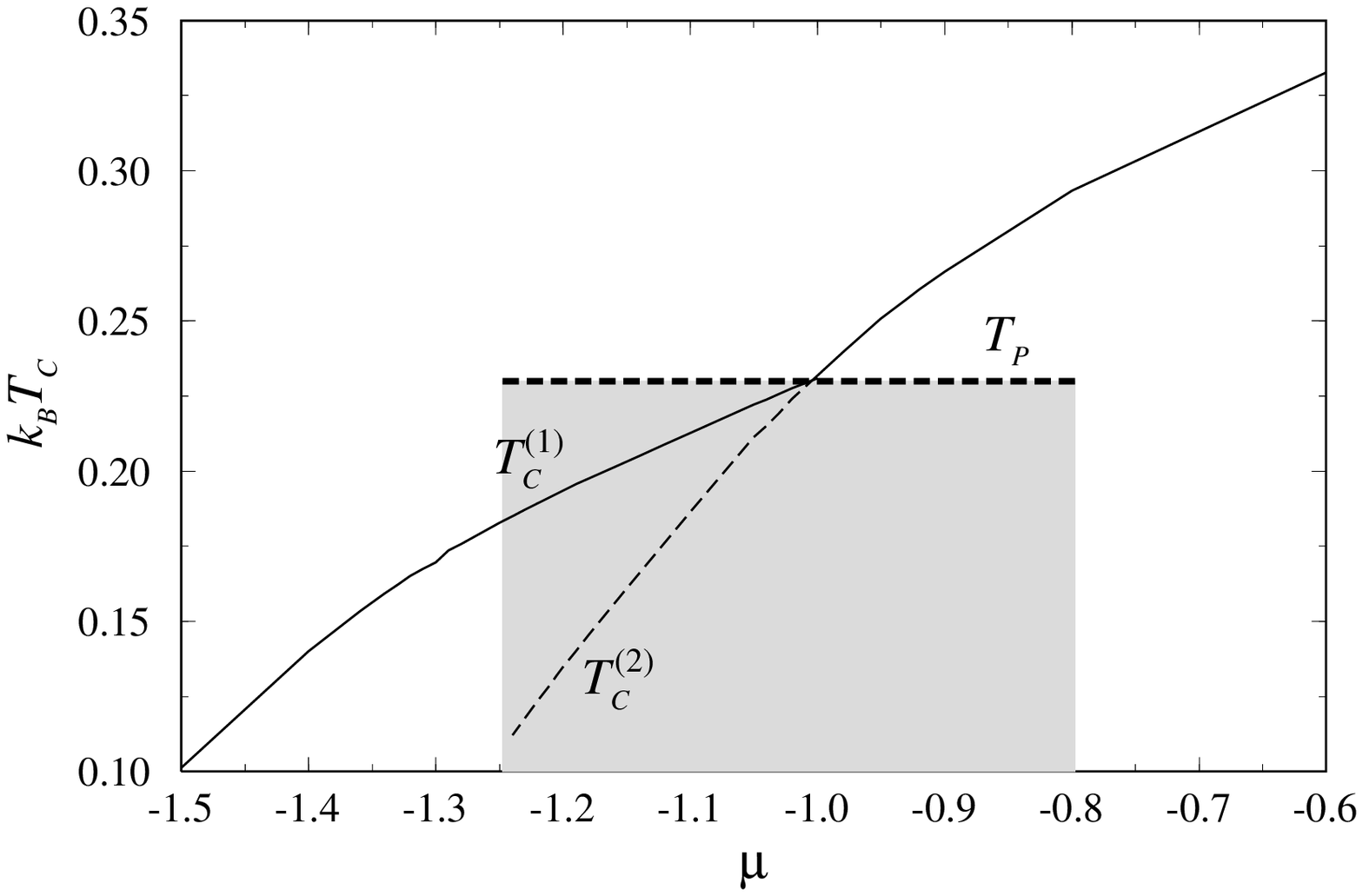}
\caption{Main transition temperatures for the degenerate double exchange model
         (in units of $dd\sigma$-integral).
         $T_P$ is the transition temperature to the two-phase paramagnetic state.
         The shaded area shows the approximate
         range of the chemical potentials ($\mu$) when the paramagnetic state
         becomes intrinsicly inhomogeneous. $T_C^{(1)}$ and $T_C^{(2)}$ are the
         Curie temperatures for two different phases (characterized by
         lower and higher densities of the $e_g$ electrons, respectively).}
\label{fig.2bTC}
\end{figure}

  Results of these calculations are shown in Fig.~\ref{fig.2bTC}.
For $\mu$$<$$-$$1$, the magnetic transition
temperature appears to be lower than the point of PM phase separation
($T_P$).
Taking into account that $T_P$$\simeq$$0.23$ and using results of Fig.~\ref{fig.twophase},
$\mu$$<$$-$$1$
roughly corresponds to the densities $\langle \overline{n} \rangle$$<$$0.5$.
In this region, $T_C$ should be calculated independently for two
different phases. The calculations can be done using Eq.~(\ref{eqn:Tint1}).
Not surprisingly that different phases
are characterized by different $T_C$'s.
Therefore, for the degenerate DE model we expect
the existence of {\it two magnetic transition points}. With the cooling
down of the sample, the transition to the FM state takes place first in
one of the phases, characterized by lower density (the hole-rich phase).
Then, within the interval
$T_C^{(1)}$$<$$T$$<T_C^{(2)}$, the FM phase continues to coexist with
the PM phase, persisting in the hole-deficient
part of the sample. Taking into account that
$(dd\sigma)$$\approx$$0.7$eV, the difference of
two transition temperatures, which depend on $\mu$, can be evaluated as
$0$$<$$T_C^{(2)}$$-$$T_C^{(1)}$$<$$650$K.
Finally, for $T$$<$$T_C^{(1)}$ the system exhibits
the proper FM order established in both phases.

  Formally, the opposite scenario when the phase separation occurs below
the magnetic transition temperature is also possible, and according to
Fig.~\ref{fig.2bTC} may take place when $\mu$$>$$-$$1$
($\langle \overline{n} \rangle$$\geq$$0.5$).
However, the quantitative description of this situation is beyond the
small-$\lambda$ limit, considered in the present work.

\section{Concluding Remarks}
\label{sec:discussions}

  We have applied a non-local CPA approach to the problem of
orientational spin disorder
in the double exchange systems, which was supplemented by a
mean-field theory for the analysis of magnetic transition temperature.
Our CPA approach is based on the perturbation theory expansion for the
$T$-matrix with respect to fluctuations of hoppings
from the mean value of the DE
Hamiltonian specified by the matrix elements of
the (non-local) self energy, so that in the
lowest (first) order it automatically recovers the main results of
the DE theory by de Gennes (the bandwidth narrowing in the PM state,
expression for the Curie temperature, etc.).\cite{deGennes}
Our main focus was on the correction of this theory
by higher-order effects with respect to the fluctuations, which were
included up to the second order and treated in the real space.

  In the one-orbital case,
it led to a substantial reduction of $T_C$ (up to 20\%).
Nevertheless, the obtained value of $T_C$ was
largely
overestimated
in comparison with results of
Monte Carlo calculations,\cite{Motome}
due to limitations inherent to the
mean-field approach.
Therefore, a sensible description of
spatial spin correlations, beyond the mean-field approximation,
presents a very important direction
for the improvement of our model.

  It appeared, however, that
even on the level of mean-field theories
the situation is far from being
fully investigated, especially if one takes into account
the effects of orbital degeneracy and
details of realistic electronic structure for the $e_g$ electrons in the cubic
perovskite structure.
Particularly, two Van Hove
singularities at the
$(\pi,\pi,0)$ and $(0,\pi,0)$ points of the Brillouin zone may contribute to the
properties of DE systems not only in the ordered
FM state,\cite{Gorkov} but also in the case of the spin disorder,
above the magnetic transition temperature.
The singularities lead to the branching of CPA solutions for the PM state,
so that for certain values of the chemical potential the system consists
of two PM phases with two different densities.
In such a situation, the PM state becomes intrinsicly inhomogeneous,
that also determines
details of the phase transition to the FM state with the decrease of
the temperature.
The magnetic transition is characterized by the existence of two
transition points, and first takes place in one of the phases which constitutes
only a fraction of the sample.
The FM transition in the second phase takes place at yet another temperature,
which can be significantly lower than the first one.
Both magnetic transitions are continuous (of the second order).
However, they occur separately in two different phases which already
exist in the PM region, i.e. above the first magnetic transition point.\cite{comment.11}

  How is this scenario consistent with the experimental behavior of perovskite
manganites?

  It is true that there is no clear experimental evidence supporting the existence of
an individual temperature of the PM phase separation, $T_P$, and
the intrinsic inhomogeneity of the PM phase (unless it is caused by external factors
such as the chemical and structural inhomogeneities and the grain boundaries).\cite{Tomioka1}
In this sense our result can be regarded as the prediction.
On the other hand, the phase coexistence below $T_C$
is rather common, and was observed in a number of
experiments.\cite{Uehara,Fath,Lynn}
In addition,
our result naturally explains the appearance of several magnetic transition
points in perovskite manganites.
Yet, a clear difference is that experimentally
one (or sometimes both) magnetic transitions are antiferromagnetic
(typically either to the A- or CE-type AFM state), and the
situation when two consecutive transitions go to the FM state is not realized
in practice.\cite{Tokura}
Presumably, the difference is caused by the limitation of our analysis
by the PM and FM states, while in reality for
$\langle \overline{n} \rangle$$<$$0.7$ (i.e. when several magnetic
transition points are expected) the formation of the
(A-, CE-, and C-type)
AFM structures seems to be more natural.
From this point of view, results of our work present mainly an academical
interest: we pointed out at the principal possibility of the existence of
several magnetic transition temperatures for the DE systems, however the type of the
ordered state considered at the low temperature was not sufficiently
general.\cite{comment.12}

  Finally, we would like to discuss briefly some possible extensions of our model.

\begin{enumerate}
\item
As it was already mentioned, any realistic description of the
phase diagram of perovskite manganites would be incomplete without
the (A-, CE-, and C-type) AFM structures.
The minimal model which captures the behavior of doped manganites at
the low temperature is the double exchange combined with the isotropic
AFM SE interaction between the localized spins ($J^S$).
So, the magnetic phase diagram at $T$$=$$0$ can be understood in
terms of the anisotropy of interatomic DE interactions, caused by the anisotropy
of magnetic ordering and operating in the background of isotropic AFM
SE interactions.\cite{Springer02}
In this picture, the SE interactions were needed to shift
the reference point simultaneously in all bonds in the direction of
the
AFM coupling, while the variety of the phase diagram itself
is described by the anisotropy of the DE interactions.
However, similar shift is not applicable
for the magnetic transition temperature, or may not be the major
effect of the SE interactions.
For example, the shift of the Curie temperature by $J^S$,
$T_C$$\rightarrow$$T_C$$-$$2|J^S|$,\cite{comment.3}
will also affect the mesh of Matsubara poles
in Eq.~(\ref{eqn:Tint1}).
It will require the careful
analysis of the topology of CPA solutions near the real axis, which will
present the main obstacle for such calculations.
\item
Our main results regarding the nontrivial topology of the CPA solutions for the
PM state have been based on the second order of the
perturbation theory expansion for
the $T$-matrix.
Since the effect was so dramatic, it naturally rises the question about
importance of the higher-order terms.
It seems to be a very important problem for the future analysis.
\item
The main effect discussed in this work is based on the existence of Van Hove
singularities in the density of states of degenerate DE model.
However, the exact position of these singularities, or even the fact of
their existence in realistic compounds depend on many other factors, such as the
Mn($3d$)-O($2p$) hybridization, the cation and structural disorder, the
purity of sample, etc.
All these factors may significantly alter conclusions of our work, if one
try to apply them to realistic compounds.
\end{enumerate}

\section*{Acknowledgments}
  I thank F.~Aryasetiawan for discussion of the energy integration around the
branch-point in the complex plane; Y.~Tomioka - for discussion of the experimental
behavior of manganites, and A.~I.~Liechtenstein -
for discussion of causality problems in non-local CPA.
The present work is partly supported
by New Energy and Industrial Technology
Development Organization (NEDO).

\appendix
\section{Orientational Average of
the
$T$-matrix: \\
One-Orbital Model}
\label{apndx:1}

  After averaging over the directions of magnetic moments
with the distribution function given by Eq.~(\ref{eqn:dfunction}),
different contributions to the $T$-matrix (see Table~\ref{tab:terms})
become:
$$
\overline{\Delta {\cal H}_{\bf 00}}=-\sigma_0,
$$
$$
\overline{\Delta {\cal H}_{\bf 00}\overline{G}_{\bf 00}\Delta {\cal H}_{\bf 00}}=\sigma_0^2 g_0,
$$
$$
\overline{\sum_{\bf i}\{\Delta {\cal H}_{\bf 00}\overline{G}_{\bf 0i}\Delta {\cal H}_{\bf i0}\}}=
\overline{\sum_{\bf i}\{\Delta {\cal H}_{\bf 0i}\overline{G}_{\bf i0}\Delta {\cal H}_{\bf 00}\}}=
-6 \sigma_0 \sigma_1 g_1
+\frac{4}{15}\sigma_0  g_1 \lambda^2,
$$
$$
\overline{\sum_{\bf i}\{\Delta {\cal H}_{\bf 0i}\overline{G}_{\bf ii}\Delta {\cal H}_{\bf i0}\}}=
\left(6 \sigma_1^2 + \frac{1}{3}\right) g_0 -
\frac{2}{15}\left(4 \sigma_1^2 + \frac{1}{6}\right) g_0 \lambda^2,
$$
$$
\overline{\Delta {\cal H}_{\bf 01}}= \sigma_1 - \frac{2}{45} \lambda^2,
$$
$$
\overline{\Delta {\cal H}_{\bf 00}\overline{G}_{\bf 01}\Delta {\cal H}_{\bf 11}}= \sigma_0^2 g_1,
$$
$$
\overline{\Delta {\cal H}_{\bf 00}\overline{G}_{\bf 00}\Delta {\cal H}_{\bf 01}}=
\overline{\Delta {\cal H}_{\bf 01}\overline{G}_{\bf 11}\Delta {\cal H}_{\bf 11}}=
-\sigma_0 \sigma_1 g_0
+ \frac{2}{45} \sigma_0 g_0 \lambda^2,
$$
$$
\overline{\sum_{\bf i}\{\Delta {\cal H}_{\bf 0i}\overline{G}_{\bf i0}\Delta {\cal H}_{\bf 01}\}}=
5 \sigma_1^2 g_1  - \frac{4}{9} \sigma_1 g_1 \lambda^2,
$$
$$
\overline{\sum_{\bf i}\{\Delta {\cal H}_{\bf 01}\overline{G}_{\bf 1i}\Delta {\cal H}_{\bf i1}\}}=
5\left( \sigma_1^2 + \frac{2}{9}\sigma_1 + \frac{1}{27} \right) g_1
-\frac{14}{27} \left( \sigma_1 - \frac{1}{3} \right) g_1 \lambda^2,
$$
$$
\overline{\Delta {\cal H}_{\bf 01}\overline{G}_{\bf 10}\Delta {\cal H}_{\bf 01}}=
\left( \sigma_1^2 + \frac{1}{18} \right) g_1
-\frac{1}{45} \left(4 \sigma_1 + \frac{1}{6} \right) g_1 \lambda^2,
$$
and
$$
\overline{\sum_{\bf ij}\{\Delta {\cal H}_{\bf 0i}\overline{G}_{\bf ij}\Delta {\cal H}_{\bf j1}\}}=
4 \sigma_1 \left(\sigma_1 + \frac{2}{9} \right) g_1 -
\frac{8}{135} \left( 7 \sigma_1 + \frac{2}{3} \right) g_1 \lambda^2.
$$

\section{Orientational Average of
the
$T$-matrix: \\
the Case of Two \lowercase{$e_g$} Orbitals}
\label{apndx:2}

  Due to degeneracy,
it is sufficient to consider
the contributions to only one site-diagonal element of the
averaged $T$-matrix, say
$\overline{T}_{\bf 00}^{11}(z)$. Results of the orientational
averaging for $\overline{T}_{\bf 00}^{22}(z)$
will be identical.
Then, for different contributions listed in Table~\ref{tab:terms}
we have:
$$
[\overline{\Delta {\cal H}_{\bf 00}}]^{11}=-\sigma_0,
$$
$$
[\overline{\Delta {\cal H}_{\bf 00}\overline{G}_{\bf 00}\Delta {\cal H}_{\bf 00}}]^{11}
=\sigma_0^2 g_0,
$$
$$
[\overline{\sum_{\bf i}\{\Delta {\cal H}_{\bf 00}\overline{G}_{\bf 0i}\Delta {\cal H}_{\bf i0}\}}]^{11}=
[\overline{\sum_{\bf i}\{\Delta {\cal H}_{\bf 0i}\overline{G}_{\bf i0}\Delta {\cal H}_{\bf 00}\}}]^{11}=
-3 \sigma_0 \left( \sigma_1 g_1 + \sigma_2 g_2 \right)
+\frac{2}{15}\sigma_0  g_2 \lambda^2,
$$
and
$$
[\overline{\sum_{\bf i}\{\Delta {\cal H}_{\bf 0i}\overline{G}_{\bf ii}\Delta {\cal H}_{\bf i0}\}}]^{11}=
3 \left( \sigma_1^2 + \sigma_2^2 + \frac{1}{18}\right) g_0 -
\frac{1}{15}\left(4 \sigma_2^2 + \frac{1}{6}\right) g_0 \lambda^2.
$$

  For the bond ${\bf 0}$-${\bf 1}$,
results of the orientational averaging of the diagonal $11$ and $22$ matrix elements
will be different. The contributions to $\overline{T}_{\bf 01}^{11}(z)$
are given by:
$$
[\overline{\Delta {\cal H}_{\bf 01}}]^{11}= \sigma_1,
$$
$$
[\overline{\Delta {\cal H}_{\bf 00}\overline{G}_{\bf 01}\Delta {\cal H}_{\bf 11}}]^{11}=
\sigma_0^2 g_1,
$$
$$
[\overline{\Delta {\cal H}_{\bf 00}\overline{G}_{\bf 00}\Delta {\cal H}_{\bf 01}}]^{11}=
[\overline{\Delta {\cal H}_{\bf 01}\overline{G}_{\bf 11}\Delta {\cal H}_{\bf 11}}]^{11}=
-\sigma_0 \sigma_1 g_0,
$$
$$
[\overline{\sum_{\bf i}\{\Delta {\cal H}_{\bf 0i}\overline{G}_{\bf i0}\Delta {\cal H}_{\bf 01}\}}]^{11}=
2 \sigma_1^2 g_1 + 3 \sigma_1 \sigma_2 g_2
- \frac{2}{15} \sigma_1 g_2 \lambda^2,
$$
$$
[\overline{\sum_{\bf i}\{\Delta {\cal H}_{\bf 01}\overline{G}_{\bf 1i}\Delta {\cal H}_{\bf i1}\}}]^{11}=
2 \sigma_1^2 g_1 + \sigma_1 \left( 3 \sigma_2 + \frac{2}{3} \right) g_2
- \frac{8}{45} \sigma_1 g_2 \lambda^2,
$$
$$
[\overline{\Delta {\cal H}_{\bf 01}\overline{G}_{\bf 10}\Delta {\cal H}_{\bf 01}}]^{11}=
\sigma_1^2 g_1,
$$
and
\begin{eqnarray*}
  [\overline{\sum_{\bf ij}\{\Delta {\cal H}_{\bf 0i}\overline{G}_{\bf ij}\Delta {\cal H}_{\bf j1}\}}]^{11} =
\\
  \frac{1}{4} \left\{ \left( \sigma_1^2 + 6 \sigma_1 \sigma_2 +9 \sigma_2^2
+\frac{2}{3} \sigma_1 + 2 \sigma_2 \right) g_1 +
3 \left( \sigma_1^2 -2 \sigma_1 \sigma_2 + \sigma_2^2 -\frac{2}{9} \sigma_1 + \frac{2}{9} \sigma_2
\right) g_2
\right\} -
\\
  \frac{1}{90} \left\{ \left( \frac{}{} 7 \sigma_1 +21 \sigma_2  + 2 \right) g_1 -
\left( 7 \sigma_1 -7 \sigma_2  - \frac{2}{3} \right) g_2
\right\} \lambda^2;
\end{eqnarray*}
while similar contributions to $\overline{T}_{\bf 01}^{22}(z)$ are given by:
$$
[\overline{\Delta {\cal H}_{\bf 01}}]^{22}= \sigma_2 - \frac{2}{45} \lambda^2,
$$
$$
[\overline{\Delta {\cal H}_{\bf 00}\overline{G}_{\bf 01}\Delta {\cal H}_{\bf 11}}]^{22}=
\sigma_0^2 g_2,
$$
$$
[\overline{\Delta {\cal H}_{\bf 00}\overline{G}_{\bf 00}\Delta {\cal H}_{\bf 01}}]^{22}=
[\overline{\Delta {\cal H}_{\bf 01}\overline{G}_{\bf 11}\Delta {\cal H}_{\bf 11}}]^{22}=
-\sigma_0 \sigma_2 g_0 + \frac{2}{45} \sigma_0 g_0 \lambda^2,
$$
$$
[\overline{\sum_{\bf i}\{\Delta {\cal H}_{\bf 0i}\overline{G}_{\bf i0}\Delta {\cal H}_{\bf 01}\}}]^{22}=
3 \sigma_1 \sigma_2 g_1 + 2 \sigma_2^2 g_2
- \frac{2}{15} \left( \sigma_1 g_1 +\frac{4}{3} \sigma_2 g_2 \right) \lambda^2,
$$
$$
[\overline{\sum_{\bf i}\{\Delta {\cal H}_{\bf 01}\overline{G}_{\bf 1i}\Delta {\cal H}_{\bf i1}\}}]^{22}=
3 \sigma_1 \sigma_2 g_1 + 2 \left( \sigma_2^2 + \frac{2}{9} \sigma_2 + \frac{1}{27} \right) g_2
- \frac{2}{15} \left\{ \sigma_1 g_1 +\frac{2}{9} \left( 7 \sigma_2
+\frac{2}{3} \right) g_2 \right\} \lambda^2,
$$
$$
[\overline{\Delta {\cal H}_{\bf 01}\overline{G}_{\bf 10}\Delta {\cal H}_{\bf 01}}]^{22}=
\left( \sigma_2^2 +\frac{1}{18} \right) g_2 -\frac{1}{45}
\left( 4 \sigma_2 + \frac{1}{6} \right) g_2 \lambda^2,
$$
and
\begin{eqnarray*}
  [\overline{\sum_{\bf ij}\{\Delta {\cal H}_{\bf 0i}\overline{G}_{\bf ij}\Delta {\cal H}_{\bf j1}\}}]^{22}  =
\\
  \frac{1}{4} \left\{ 3 \left( \sigma_1^2 - 2 \sigma_1 \sigma_2 + \sigma_2^2
-\frac{2}{9} \sigma_1 + \frac{2}{9} \sigma_2 \right) g_1 +
 \left( 9 \sigma_1^2 +6 \sigma_1 \sigma_2 + \sigma_2^2 +\frac{2}{3} \sigma_1 + \frac{2}{9} \sigma_2
\right) g_2
\right\} +
\\
  \frac{1}{90} \left\{ \left( 7 \sigma_1 -7 \sigma_2  - \frac{2}{3} \right) g_1 -
\left( 7 \sigma_1 +\frac{7}{3} \sigma_2  + \frac{2}{9} \right) g_2
\right\} \lambda^2.
\end{eqnarray*}

\section{Change of the integrated density of states in the
double-valued region}
\label{apndx:3}

  In this appendix we discuss some practical aspects of calculations of the
difference $\overline{n}^{(2)}(z)$$-$$\overline{n}^{(1)}(z)$ between two
CPA solutions in the PM states. According to Ducastelle,\cite{Ducastelle}
$\overline{n}(z)$ is given by the following expression:
\begin{equation}
\overline{n}(z) = \frac{1}{\pi N} {\rm Im}{\rm Tr}
\left\{ \ln \widehat{\overline{G}}(z) -
\overline{
\ln \left(\widehat{1} - \left[ \widehat{\cal H} -
\widehat{\overline{\cal H}}(z) \right] \widehat{\overline{G}}(z) \right)
}
\right\},
\label{eqn:na_in_CPA}
\end{equation}
where ${\rm Tr}$ is the trace over site and orbital indices, $N$ is the number
of atomic sites, and the hat-symbols here stand for the matrices in the orbital and
atomic coordinates space.

  Let us begin with the first term. In the second order of
$\widehat{\overline{G}}\mbox{$^{(2)}$}(z)$$-$$\widehat{\overline{G}}\mbox{$^{(1)}$}(z)$
we have:
\begin{eqnarray*}
\ln \widehat{\overline{G}}\mbox{$^{(2)}$}(z) - \ln \widehat{\overline{G}}\mbox{$^{(1)}$}(z) & \simeq &
2 \left\{ \widehat{\overline{G}}\mbox{$^{(2)}$}(z) - \widehat{\overline{G}}\mbox{$^{(1)}$}(z) \right\}
\left\{ \widehat{\overline{G}}\mbox{$^{(2)}$}(z) + \widehat{\overline{G}}\mbox{$^{(1)}$}(z) \right\}^{-1} \\
  & = & 2 \left\{ [\widehat{\overline{G}}\mbox{$^{(1)}$}(z)]^{-1} -
[\widehat{\overline{G}}\mbox{$^{(2)}$}(z)]^{-1} \right\}
\left\{ [\widehat{\overline{G}}\mbox{$^{(1)}$}(z)]^{-1} +
[\widehat{\overline{G}}\mbox{$^{(2)}$}(z)]^{-1} \right\}^{-1},
\end{eqnarray*}
which can be further transformed using the definition of the Green function
(\ref{eqn:disorder_G}) as
$$
\ln \widehat{\overline{G}}\mbox{$^{(2)}$}(z) - \ln \widehat{\overline{G}}\mbox{$^{(1)}$}(z) \simeq
2 \left\{ \widehat{\overline{\cal H}}\mbox{$^{(2)}$}(z) -
\widehat{\overline{\cal H}}\mbox{$^{(1)}$}(z) \right\}
\left\{ 2z - \widehat{\overline{\cal H}}\mbox{$^{(1)}$}(z) - \widehat{\overline{\cal H}}\mbox{$^{(2)}$}(z)
\right\}^{-1}.
$$
The inverse matrix $\left\{...\right\}^{-1}$ can be calculated in the same way as the
Green function (\ref{eqn:disorder_G}). Then, if $R_0(z)$ is the site-diagonal element of
$\left\{...\right\}^{-1}$, and $R_1(z)$ and $R_2(z)$ are the site-off-diagonal ones
corresponding to the $x^2$$-$$y^2$ and $3z^2$$-$$r^2$ states for the bond ${\bf 0}$-${\bf 1}$,
we can write
$$
\frac{1}{N} {\rm Tr} \left\{
\ln \widehat{\overline{G}}\mbox{$^{(2)}$}(z) - \ln \widehat{\overline{G}}\mbox{$^{(1)}$}(z)
\right\} \simeq
4 \Delta \sigma_0(z) R_0(z) - 12 \left\{ \Delta \sigma_1(z) R_1(z) + \Delta \sigma_2(z) R_2(z) \right\},
$$
where $\Delta \sigma_\alpha (z)$$=$$\sigma^{(2)}_\alpha (z)$$-$$\sigma^{(1)}_\alpha (z)$.

  In the second term of Eq.~(\ref{eqn:na_in_CPA}) (the so-called vortex correction)
we expand $\ln$ up to the second order of
$[ \widehat{\cal H} - \widehat{\overline{\cal H}}(z) ]$. This expansion is
necessary to preserve the variational properties of our CPA formalism,\cite{Ducastelle}
which is based on the same approximation (\ref{eqn:Tsecond}) for the $T$-matrix. Thus,
we have
$$
\ln \left(\widehat{1} - \left[ \widehat{\cal H} -
\widehat{\overline{\cal H}}(z) \right] \widehat{\overline{G}}(z) \right) \simeq -
\left\{ \left[ \widehat{\cal H} - \widehat{\overline{\cal H}}(z) \right] + \frac{1}{2}
 \left[ \widehat{\cal H} - \widehat{\overline{\cal H}}(z) \right]
\widehat{\overline{G}}(z)
\left[ \widehat{\cal H} - \widehat{\overline{\cal H}}(z) \right] \right\}
\widehat{\overline{G}}(z).
$$
In order to calculate the thermal average of this expression, we note that
$$
\overline{
\left[ \widehat{\cal H} - \widehat{\overline{\cal H}}(z) \right]
\widehat{\overline{G}}(z)
\left[ \widehat{\cal H} - \widehat{\overline{\cal H}}(z) \right] }
= - \overline{ \left[ \widehat{\cal H} - \widehat{\overline{\cal H}}(z) \right] },
$$
which immediately follows from the CPA equations
(\ref{eqn:CPAconditions}) under condition
(\ref{eqn:Tsecond}). Thus,
$$
\overline{
\ln \left(\widehat{1} - \left[ \widehat{\cal H} -
\widehat{\overline{\cal H}}(z) \right] \widehat{\overline{G}}(z) \right) }
\simeq -\frac{1}{2}
\overline{
\left[ \widehat{\cal H} - \widehat{\overline{\cal H}}(z) \right]
\widehat{\overline{G}}(z) },
$$
and corresponding contribution to the integrated density of states is given by
$$
\frac{1}{N} {\rm Tr}~
\overline{
\ln \left(\widehat{1} - \left[ \widehat{\cal H} -
\widehat{\overline{\cal H}}(z) \right] \widehat{\overline{G}}(z) \right) } \simeq
\sigma_0(z) g_0(z) -3\left\{ \sigma_1(z) g_1(z) + \sigma_2(z) g_2(z) \right\}.
$$


\begin{table}
\caption{Different contributions to the site-diagonal ($T_{\bf 00}$)
         and site-off-diagonal elements ($T_{\bf 01}$) of the
         $T$-matrix in the second order with respect to the fluctuations
         $\Delta {\cal H}_{\bf ij}$$=$${\cal H}_{\bf ij}$$-$$\overline{\cal H}_{\bf ij}$
         in the real space.
         The column 'comment' is used to explain the position of intermediate sites
         used in the summation in Fig.~\protect\ref{fig.cluster}.}
\label{tab:terms}
\begin{center}
\begin{tabular}{lc}
contribution                 &   comment    \\
\tableline
~~~~~\underline{element $T_{\bf 00}$}:                        &              \\
~~~~~$\Delta {\cal H}_{\bf 00}$ &                \\
~~~~~$\Delta {\cal H}_{\bf 00}\overline{G}_{\bf 00}\Delta {\cal H}_{\bf 00}$
               &              \\
$\sum_{\bf i}\{\Delta {\cal H}_{\bf 00}\overline{G}_{\bf 0i}\Delta {\cal H}_{\bf i0}\}$
                & ${\bf i}=$ 1-6  \\
$\sum_{\bf i}\{\Delta {\cal H}_{\bf 0i}\overline{G}_{\bf i0}\Delta {\cal H}_{\bf 00}\}$
                & ${\bf i}=$ 1-6  \\
$\sum_{\bf i}\{\Delta {\cal H}_{\bf 0i}\overline{G}_{\bf ii}\Delta {\cal H}_{\bf i0}\}$
                & ${\bf i}=$ 1-6  \\
~~~~~\underline{element $T_{\bf 01}$}:                        &              \\
~~~~~$\Delta {\cal H}_{\bf 01}$             &              \\
~~~~~$\Delta {\cal H}_{\bf 00}\overline{G}_{\bf 01}\Delta {\cal H}_{\bf 11}$
                &              \\
~~~~~$\Delta {\cal H}_{\bf 00}\overline{G}_{\bf 00}\Delta {\cal H}_{\bf 01}$
                &              \\
~~~~~$\Delta {\cal H}_{\bf 01}\overline{G}_{\bf 11}\Delta {\cal H}_{\bf 11}$
                &              \\
$\sum_{\bf i}\{\Delta {\cal H}_{\bf 0i}\overline{G}_{\bf i0}\Delta {\cal H}_{\bf 01}\}$
                & ${\bf i}=$ 1-5  \\
$\sum_{\bf i}\{\Delta {\cal H}_{\bf 01}\overline{G}_{\bf 1i}\Delta {\cal H}_{\bf i1}\}$
                & ${\bf i}=$ 7-11 \\
~~~~~$\Delta {\cal H}_{\bf 01}\overline{G}_{\bf 10}\Delta {\cal H}_{\bf 01}$
                &                 \\
$\sum_{\bf ij}\{\Delta {\cal H}_{\bf 0i}\overline{G}_{\bf ij}\Delta {\cal H}_{\bf j1}\}$
                & (${\bf i}$,${\bf j}$) = (3,7), (4,8), (5,9), (6,10) \\
\end{tabular}
\end{center}
\end{table}

\begin{thebibliography}{10}
\bibitem{comment.6}
Perhaps the absolute value of $T_C$
is not a serious problem as it can be easily adjusted by adding a phenomenological
AFM SE interaction between the localized
spins.\cite{Alonso,Dzero}
The situation with the doping-dependence appears to be more serious. The experimental
$T_C$ is not monotonous and has a maximum around $x$$\approx$$0.3$
(in La$_{1-x}$Sr$_x$MnO$_3$).\cite{Urushibara}
The maximum does exist in the one-orbital DE model
(at $x$$=$$0.5$).\cite{Furukawa} However, it is only an artifact of the
one-orbital description, whereas in a more realistic DE model $T_C$
decreases as the function of $x$.\cite{Dzero}
\bibitem{comment.13}
Formally, the correlations can be included in a systematic way by adding corresponding
terms to the trial Hamiltonian, which specifies the form of the distribution
function (\ref{eqn:dfunction0}).\cite{DLM,Tyablikov,Uhl}
However, it makes the calculations costly, and the main idea -- less
transparent. Therefore, at the first stage we neglect the correlations.
\bibitem{comment.4}
Note for example that the straightforward averaging of the Hamiltonian
(\ref{eqn:DE_Hamiltonian}) with the distribution functions
(\ref{eqn:dfunction0}) will lead to an unphysical result:
since $\cos \frac{\theta}{2}$ is not symmetric function of
$\cos \theta$, the self energy and the electron free energy will
include odd contributions of $\lambda$.
\bibitem{comment.8}
The factor '6' in the second term of Eq.~(\ref{eqn:Lloyd})
stands for the coordination number
and the sign '$-$' is related with the definition of the non-local part of
the self energy -- see Eq.~(\ref{eqn:disorder_H}).
\bibitem{comment.9}
The energy integration in Eq.~(\ref{eqn:DvT}) is rather straightforward. For example,
the integral along the real axis can be replaced by a contour integral in the
upper half plane plus contributions coming from residues calculated at Matsubara
poles $z_n$$=$$\mu$$+$$i\pi k_BT(2n$$+$$1)$, $n$$=$ 0, 1, 2, ...\cite{Wildberger}
\bibitem{comment.10}
In fact, different authors reported somewhat different estimates for $T_C$ based
on the DMFT. The value $0.0809$ reported in Ref.~\onlinecite{Motome}
for $\overline{n}$$=$$0.5$ in units of the half bandwidth $\frac{1}{2}W$
would roughly correspond to $T_C$$\simeq$$0.49t_0$. On the other hand,
the values $(E_D/W,T_C/E_D)$$=$$(0.106,0.21)$ and $(0.081,0.19)$ reported
in Ref.~\onlinecite{Millis} for $\overline{n}$$=$$0.5$ and $\overline{n}$$=$$0.25$
would yield  $T_C$$\simeq$$0.27t_0$ and $0.18t_0$, respectively.
In any case, all these values are higher than the ones obtained in the
non-local CPA approach in the present work.
\bibitem{comment.3}
Note, that by combining the DE model with the phenomenological
SE interaction between nearest neighbors in the form
$E_S$$=$$-$$\frac{1}{2}|J^S|\sum_{\bf ij} {\bf e_i} \cdot {\bf e_j}$,
the Curie temperature can be reduced, in the mean-field approximation,
as $T_C$$\rightarrow$$T_C$$-$$2|J^S|$. In three-dimensional manganites
$R_{1-x}D_x$MnO$_3$, the exchange constant can be estimated
(from the low-temperature spin-wave dispersion) as
$|J^S|$$\sim$$60$-$70$ meV.\cite{Springer02}
\bibitem{comment.14}
In these calculations, we start
CPA iterations at the point
${\rm Re}(z)$$=$${\rm arbitrary}$, ${\rm Im}(z)$$=$$0.5$ with two
self-consistent solutions corresponding to one of such
points, say $z$$=$$($$-$$1.1,0.5)$, which were obtained in advance, and move along
the imaginary axis until the merging of two solutions.
\bibitem{comment.15}
In practical terms, we start the iterations at $z_0$ with two different sets for the
self energy, which guarantee two different CPA solutions, and then at each
consecutive pole $z_n$ start with
the self-consistent CPA solutions obtained
for $z_{n-1}$ (and so on until merging of the two solutions).
\bibitem{comment.2}
Eq.~(\ref{eqn:TCfunc})
is obtained by using the Lloyd formula
for the integrated density of states and is nothing but the generalization
of Eq.~(\ref{eqn:Lloyd}) for the two-orbital case, which includes also the
trace over the orbital indices. The latter is responsible for the factor '2'
in the first term and gives rise to the two contributions in the second term
of Eq.~(\ref{eqn:TCfunc}).
\bibitem{comment.11}
In this sense, our scenario is qualitatively different from the FM-PM
phase separation encountered in the one-orbital DE model with an additional
AFM SE interaction, which typically accompanies the discontinuous
(first-order) phase transition to the FM state.\cite{Alonso,comment.5}
\bibitem{comment.12}
The choice of the FM phase
was general from the viewpoint of the pure DE model, where the FM ground
state at $T$$=$$0$ is the only possibility.
However, the assignment was not sufficiently general from the viewpoint
of possible modifications of the DE model, including the AFM
SE interactions between the localized spins, for example.
\end{thebibliography}
\end{document}